\pdfoutput=1
\documentclass[runningheads,rgb]{llncs}
\usepackage[T1]{fontenc}
\usepackage[utf8]{inputenc}
\usepackage{mathtools,amsfonts,amssymb}
\usepackage{nicefrac}
\usepackage{graphicx}
\usepackage[table]{xcolor}
\usepackage{listings}
\usepackage{csquotes}
\usepackage{tikz}
\usepackage[binary-units=true]{siunitx}
\usepackage[textsize=scriptsize, disable]{todonotes}
\usepackage{subcaption}
\usepackage{hyperref}
\usepackage{xfrac}
\usepackage{stmaryrd}
\usepackage{bm}
\usepackage{etoolbox}

\makeatletter
\let\oldtheequation\theequation
\renewcommand\tagform@[1]{\maketag@@@{\ignorespaces#1\unskip\@@italiccorr}}
\renewcommand\theequation{(\oldtheequation)}
\makeatother

\usetikzlibrary{arrows.meta, calc, positioning, matrix, shapes.geometric, quantikz, backgrounds, chains, shapes, shapes.arrows, graphs}

\tikzset{%
	font={\footnotesize},
	vertex/.style={draw,circle,inner sep=0pt,minimum width=0.5cm,minimum height=0.5cm},
	zeroterm/.style={below,inner sep=0pt,font=\tiny}
}

\newcommand{\qop}[1]{\ensuremath{\mathit{#1}}}

\usepackage[style=ieee, maxnames=6, minnames=1, date=year, doi=false,isbn=false,url=false,backend=biber, sortcites, dashed=false]{biblatex}
\csundef{example}
\csundef{c@example}
\spnewtheorem{example}[theorem]{Example}{\bfseries}{\em}

\definecolor{omega0}{RGB}{191,64,64}  
\definecolor{omega1}{RGB}{191,159,64} 
\definecolor{omega2}{RGB}{128,191,64} 
\definecolor{omega3}{RGB}{64,191,96}  
\definecolor{omega4}{RGB}{64,191,191} 
\definecolor{omega5}{RGB}{64,96,191}  
\definecolor{omega6}{RGB}{128,64,191} 
\definecolor{omega7}{RGB}{191,64,159} 

\tikzset{%
	terminal/.style={draw,rectangle,inner sep=2pt,font=\footnotesize,very thick},
	zeronode/.style={fill, draw, circle, minimum width=2pt, inner sep=0pt,color=black},
	qubit/.style={draw,circle,inner sep=0pt,minimum width=0.4cm,minimum height=0.4cm,font=\footnotesize,color=black, thin},
	edgeOne/.style={color=omega0,ultra thick},
	edgeMOne/.style={color=omega4,ultra thick},
	edgeSqrt/.style={color=omega0},
	edgeMSqrt/.style={color=omega4},
	edgeSqrt3/.style={color=omega0,thin},
	edgeOmega0/.style={color=omega0,ultra thick},
	edgeOmega1/.style={color=omega1,ultra thick},
	edgeOmega2/.style={color=omega2,ultra thick},
	edgeOmega3/.style={color=omega3,ultra thick},
	edgeOmega4/.style={color=omega4,ultra thick},
	edgeOmega5/.style={color=omega5,ultra thick},
	edgeOmega6/.style={color=omega6,ultra thick},
	edgeOmega7/.style={color=omega7,ultra thick},
	edge0/.style={edge from parent path={(\tikzparentnode) to[out=-130, in=90] (\tikzchildnode)}},
	edge1/.style={edge from parent path={(\tikzparentnode) to[out=-100, in=90] (\tikzchildnode)}},
	edge2/.style={edge from parent path={(\tikzparentnode) to[out=-80, in=90] (\tikzchildnode)}},
	edge3/.style={edge from parent path={(\tikzparentnode) to[out=-50, in=90] (\tikzchildnode)}},
}

\newsavebox{\legend}
\sbox{\legend}{%
	\begin{tikzpicture}
	\draw[edgeOmega7]  (0, 0) node[left] {$\omega^7$}  -- (1, 0);
	\draw[edgeOmega6]  (0, 0.5) node[left] {$\omega^6$}  -- (1, 0.5);
	\draw[edgeOmega5]  (0, 1) node[left] {$\omega^5$}  -- (1, 1);
	\draw[edgeOmega4]  (0, 1.5) node[left] {$\omega^4$}  -- (1, 1.5);
	\draw[edgeOmega3]  (0, 2) node[left] {$\omega^3$}  -- (1, 2);
	\draw[edgeOmega2]  (0, 2.5) node[left] {$\omega^2$}  -- (1, 2.5);
	\draw[edgeOmega1]  (0, 3) node[left,overlay] {$\frac{(1+i)}{\sqrt{2}} = \omega^{\phantom{2}}$}  -- (1, 3);
	\draw[edgeOmega0]  (0, 3.5) node[left] {$1$}  -- (1, 3.5);
	\draw[edgeSqrt3]   (0, 4) node[left] {$\frac{1}{\sqrt{8}}$} -- (1, 4);
	\end{tikzpicture}
}

\newsavebox{\qftdd}
\sbox{\qftdd}{%
	\begin{tikzpicture}[level distance = 1cm,
	level 1/.style={sibling distance=2cm},
	level 2/.style={sibling distance=0.5cm}]
		\node[qubit] (q2) {$2$}
		child{
			node[qubit] (q1a) {$1$} edge from parent[edgeOmega0, edge0]
			child {
				node[qubit] (q0a) {$0$} edge from parent[edgeOmega0, edge0]
			}
			child {
				node[qubit] (q0b) {$0$} edge from parent[edgeOmega0, edge1]
			}
			child {
				node[qubit] (q0c) {$0$} edge from parent[edgeOmega0, edge2]
			}
			child {
				node[qubit] (q0d) {$0$} edge from parent[edgeOmega4, edge3]
			}
		}
		child{
			node[qubit] (q1b) {$1$} edge from parent[edgeOmega0, edge1]
			child {
				node[qubit] (q0e) {$0$} edge from parent[edgeOmega0, edge0]
			}
			child {
				node[qubit] (q0f) {$0$} edge from parent[edgeOmega0, edge1]
			}
			child {
				node[qubit] (q0g) {$0$} edge from parent[edgeOmega0, edge2]
			}
			child {
				node[qubit] (q0h) {$0$} edge from parent[edgeOmega4, edge3]
			}	
		}
		child{
			node[qubit] (q1c) {$1$} edge from parent[edgeOmega0, edge2]		
			child {
				node[qubit] (q0i) {$0$} edge from parent[edgeOmega0, edge0]
			}
			child {
				node[qubit] (q0j) {$0$} edge from parent[edgeOmega0, edge1]
			}
			child {
				node[qubit] (q0k) {$0$} edge from parent[edgeOmega0, edge2]
			}
			child {
				node[qubit] (q0l) {$0$} edge from parent[edgeOmega4, edge3]
			}
		}
		child{
			node[qubit] (q1d) {$1$}	edge from parent[edgeOmega0, edge3]	
			child {
				node[qubit] (q0m) {$0$} edge from parent[edgeOmega0, edge0]
			}
			child {
				node[qubit] (q0n) {$0$} edge from parent[edgeOmega0, edge1]
			}
			child {
				node[qubit] (q0o) {$0$} edge from parent[edgeOmega0, edge2]
			}
			child {
				node[qubit] (q0p) {$0$} edge from parent[edgeOmega4, edge3]
			}
		};
		\node[terminal, below=4cm of q2] (t) {}; 
		
		\draw[edgeSqrt3] ($(q2)+(0,0.5cm)$) -- (q2);
		
		\draw[edgeOmega0] (q0a) to[out=-130, in=90] (t);
		\draw[edgeOmega0] (q0a) to[out=-100, in=90] (t);
		\draw[edgeOmega0] (q0a) to[out=-80, in=90] (t);
		\draw[edgeOmega1] (q0a) to[out=-50, in=90] (t);
		
		\draw[edgeOmega0] (q0b) to[out=-130, in=90] (t);
		\draw[edgeOmega0] (q0b) to[out=-100, in=90] (t);
		\draw[edgeOmega2] (q0b) to[out=-80, in=90] (t);
		\draw[edgeOmega3] (q0b) to[out=-50, in=90] (t);
		
		\draw[edgeOmega0] (q0c) to[out=-130, in=90] (t);
		\draw[edgeOmega2] (q0c) to[out=-100, in=90] (t);
		\draw[edgeOmega0] (q0c) to[out=-80, in=90] (t);
		\draw[edgeOmega2] (q0c) to[out=-50, in=90] (t);
		
		\draw[edgeOmega0] (q0d) to[out=-130, in=90] (t);
		\draw[edgeOmega2] (q0d) to[out=-100, in=90] (t);
		\draw[edgeOmega2] (q0d) to[out=-80, in=90] (t);
		\draw[edgeOmega5] (q0d) to[out=-50, in=90] (t);
		
		\draw[edgeOmega0] (q0e) to[out=-130, in=90] (t);
		\draw[edgeOmega0] (q0e) to[out=-100, in=90] (t);
		\draw[edgeOmega4] (q0e) to[out=-80, in=90] (t);
		\draw[edgeOmega5] (q0e) to[out=-50, in=90] (t);
		
		\draw[edgeOmega0] (q0f) to[out=-130, in=90] (t);
		\draw[edgeOmega0] (q0f) to[out=-100, in=90] (t);
		\draw[edgeOmega6] (q0f) to[out=-80, in=90] (t);
		\draw[edgeOmega7] (q0f) to[out=-50, in=90] (t);
		
		\draw[edgeOmega0] (q0g) to[out=-130, in=90] (t);
		\draw[edgeOmega2] (q0g) to[out=-100, in=90] (t);
		\draw[edgeOmega4] (q0g) to[out=-80, in=90] (t);
		\draw[edgeOmega7] (q0g) to[out=-50, in=90] (t);
		
		\draw[edgeOmega0] (q0h) to[out=-130, in=90] (t);
		\draw[edgeOmega2] (q0h) to[out=-100, in=90] (t);
		\draw[edgeOmega6] (q0h) to[out=-80, in=90] (t);
		\draw[edgeOmega1] (q0h) to[out=-50, in=90] (t);
		
		\draw[edgeOmega0] (q0i) to[out=-130, in=90] (t);
		\draw[edgeOmega4] (q0i) to[out=-100, in=90] (t);
		\draw[edgeOmega0] (q0i) to[out=-80, in=90] (t);
		\draw[edgeOmega5] (q0i) to[out=-50, in=90] (t);
		
		\draw[edgeOmega0] (q0j) to[out=-130, in=90] (t);
		\draw[edgeOmega4] (q0j) to[out=-100, in=90] (t);
		\draw[edgeOmega2] (q0j) to[out=-80, in=90] (t);
		\draw[edgeOmega7] (q0j) to[out=-50, in=90] (t);
		
		\draw[edgeOmega0] (q0k) to[out=-130, in=90] (t);
		\draw[edgeOmega6] (q0k) to[out=-100, in=90] (t);
		\draw[edgeOmega0] (q0k) to[out=-80, in=90] (t);
		\draw[edgeOmega7] (q0k) to[out=-50, in=90] (t);
		
		\draw[edgeOmega0] (q0l) to[out=-130, in=90] (t);
		\draw[edgeOmega6] (q0l) to[out=-100, in=90] (t);
		\draw[edgeOmega2] (q0l) to[out=-80, in=90] (t);
		\draw[edgeOmega1] (q0l) to[out=-50, in=90] (t);
		
		\draw[edgeOmega0] (q0m) to[out=-130, in=90] (t);
		\draw[edgeOmega4] (q0m) to[out=-100, in=90] (t);
		\draw[edgeOmega4] (q0m) to[out=-80, in=90] (t);
		\draw[edgeOmega1] (q0m) to[out=-50, in=90] (t);
		
		\draw[edgeOmega0] (q0n) to[out=-130, in=90] (t);
		\draw[edgeOmega4] (q0n) to[out=-100, in=90] (t);
		\draw[edgeOmega6] (q0n) to[out=-80, in=90] (t);
		\draw[edgeOmega3] (q0n) to[out=-50, in=90] (t);
		
		\draw[edgeOmega0] (q0o) to[out=-130, in=90] (t);
		\draw[edgeOmega6] (q0o) to[out=-100, in=90] (t);
		\draw[edgeOmega4] (q0o) to[out=-80, in=90] (t);
		\draw[edgeOmega3] (q0o) to[out=-50, in=90] (t);
		
		\draw[edgeOmega0] (q0p) to[out=-130, in=90] (t);
		\draw[edgeOmega6] (q0p) to[out=-100, in=90] (t);
		\draw[edgeOmega6] (q0p) to[out=-80, in=90] (t);
		\draw[edgeOmega5] (q0p) to[out=-50, in=90] (t);
			
	\end{tikzpicture}
}

 \hypersetup{pdftitle={Decision Diagrams for Quantum Computing},pdfauthor={Robert Wille, Stefan Hillmich, and Lukas Burgholzer}}
\begin{document}
\title{Decision Diagrams for Quantum Computing}
\author{Robert Wille\inst{1, 2}\orcidID{0000-0002-4993-7860}
    \and\newline\mbox{ Stefan Hillmich\inst{3}\orcidID{0000-0003-1089-3263} 
	\and Lukas Burgholzer\inst{3}\orcidID{0000-0003-4699-1316}}
}
\institute{Chair for Design Automation, Technical University of Munich, Germany \and
Software Competence Center Hagenberg GmbH (SCCH), Austria \and
Institute for Integrated Circuits, Johannes Kepler University Linz, Austria
\\\email{robert.wille@tum.de\hspace{0.5cm} stefan.hillmich@jku.at\hspace{0.5cm} lukas.burgholzer@jku.at}\\\url{https://www.cda.cit.tum.de/research/quantum/}}

\maketitle
\vspace{-2em}
\begin{abstract}
	Quantum computing promises to solve some important problems faster than conventional computations ever could.
	Currently available NISQ devices on which first practical applications are already executed demonstrate the potential---with future fault-tolerant quantum hardware for more demanding applications on the horizon.
	Nonetheless, the advantages in computing power come with challenges to be addressed in the design automation and software development community.
	In particular, non-quantum representations of states and operations, which provide the basis, e.g., for quantum circuit simulation or verification, require an exponential amount of memory.
	We propose to use decision diagrams as data structure to conquer the exponential memory requirements in many cases.
	In this chapter, we review the fundamentals on decision diagrams and highlight their applicability in the tasks of quantum circuit simulation with and without errors as well as in verification of quantum circuits. The tools presented here are all available online as open source projects.
\end{abstract}

\begingroup\small\textbf{Note:} This is a pre-print of the following chapter: Robert Wille, Stefan Hillmich, and Lukas Burgholzer, \enquote{Decision Diagrams for Quantum Computing}, published in Design Automation of Quantum Computers, edited by Rasit O. Topaloglu, 2023, Springer reproduced with permission of Springer. The final authenticated version is available online at: \url{https://doi.org/10.1007/978-3-031-15699-1_1}\endgroup

\section{Introduction}\label{sec:intro}
Quantum computing promises to speed up many important applications even in the current NISQ era~\cite{preskillQuantumComputingNISQ2018} and more so once fault-tolerance is achieved.
The underlying primitives of quantum computing are fundamentally different to conventional computations.
This introduces new challenges for design automation and software development such as the exponential memory requirement to store arbitrary quantum states and operations on non-quantum hardware.

The design automation community in the conventional domain has spent decades to successfully solve many difficult problems.
One of these solutions which especially addresses memory consumption is the usage of decision diagrams to represent information.
In the conventional domain, there exists a plethora of different types such as \emph{Binary Decision Diagrams} (BDDs,~\cite{bryantGraphbasedAlgorithmsBoolean1986}), \emph{Binary Moment Diagrams} (BMDs,~\cite{bryantVerificationArithmeticCircuits2001}), \emph{\mbox{Zero-suppressed} Decision Diagrams} (ZDDs,~\cite{minatoZerosuppressedBDDsSet1993}), or \emph{Tagged BDDs}~\cite{DBLP:conf/fmcad/DijkWM17}.
Inspired by the results achieved with decision diagram in the conventional domain, several types have been invented for the quantum domain, such as \emph{\mbox{X-decomposition} Quantum Decision Diagrams}~(XQDDs,~\cite{wangXQDDbasedVerificationMethod2008}), \emph{Quantum Decision Diagrams} (QDDs,~\cite{abdollahiAnalysisSynthesisQuantum2006}), \emph{Quantum Information Decision Diagrams} (QuIDDs,~\cite{viamontesHighperformanceQuIDDBasedSimulation2004}), or \emph{Quantum Multiple-valued Decision Diagrams} (QMDDs,~\cite{millerQMDDDecisionDiagram2006,niemannQMDDsEfficientQuantum2016}).
However, 
many researchers and engineers working in the domain of quantum computing are still rather unfamiliar with the concepts of decision diagrams and, hence, often cannot fully exploit this potential. 

In this chapter, we review decision diagrams as data structure to compactly represent quantum states and quantum operators.
To this end, we explain how decision diagrams are obtained from decomposing state vectors along with an explanation of the graphical notation.
The vector decomposition is subsequently extended to obtain decision diagrams for matrices.
Afterwards, we cover selected applications of decision diagrams for design and validation work.
More precisely, we begin by covering error-free quantum circuit simulation, which is essentially matrix-vector multiplication.
In the next step, we discuss noisy quantum circuit simulation and the advantages decision diagrams have in this application.
Additionally to simulation, we present as a main aspect of quantum circuit verification an efficient procedure to check the equivalence of quantum circuits using decision diagrams.

The remainder of this chapter is structured as follows.
\autoref{sec:dds} gives the background on decision diagrams, specifically how vectors and matrices are represented.
In \autoref{sec:sim} and \autoref{sec:noise}, we show how decision diagrams can be employed to conduct quantum circuit simulation without and with noise, respectively.
\autoref{sec:verification} explains how decision diagrams lead to more efficient equivalence checking procedures.
Finally, \autoref{sec:conclusions} concludes the chapter.

\section{Decision Diagrams}\label{sec:dds}

In this section, we describe how decision diagrams exploit redundancies in vectors and matrices to enable a compact representation in many cases.
More precisely, we first detail the representation for state vectors which we subsequently extend by a second dimension to compactly represent matrices for quantum operations.

\subsection{Representation of State Vectors}
\label{sec:vector}

The representation of a system composed of $n$ qubits on non-quantum hardware is commonly achieved through $2^n$-dimensional vector---an exponential representation.
However, a closer look at state vectors unveils that they are frequently composed of redundant entries which provide potential for a more compact representation.

\begin{example}\label{ex:dd_vector}
	Consider a quantum system with $n=3$ qubits situated in a state given by the following vector:
	\[
	 \psi = \left[\textstyle 0, 0, \frac{1}{2}, 0, \frac{1}{2}, 0, -\frac{1}{\sqrt{2}}, 0\right]^T.
	\]	
	Although exponential in size ($2^3=8$ entries), this vector only assumes three different values, namely $0$, $\frac{1}{2}$, and $-\frac{1}{\sqrt{2}}$.
\end{example}

Redundancy in the considered data can be exploited to attain a compact representation. 
To this end, we propose to employ decision diagrams. 
For conventional computations, decision diagrams such as the \emph{Binary Decision Diagram} (BDD,~\cite{bryantGraphbasedAlgorithmsBoolean1986}) are a tried and tested data structure and have been used for decades. 
For BDDs, a decomposition scheme is employed which reduces a function to be represented into corresponding sub-functions. 
Since the sub-functions usually include redundancies as well, equivalent sub-functions result which can be shared---eventually yielding a much more compact representation. 
In a similar fashion, the concept of decomposition can also be applied to represent state vectors in a more compact fashion.

Similar to decomposing a function into \mbox{sub-functions}, we decompose a given state vector with its complex entries into sub-vectors. To this end, consider a quantum system with qubits $q_{n-1}, q_1, \ldots q_0$, whereby~$q_{n-1}$ represents the most significant qubit.\footnote{The terminology \emph{most-significant qubit} refers to a position in the basis states of a quantum system and does not signify the importance of the qubit itself.} 
Then, the first $2^{n-1}$ entries of the corresponding state vector represent the amplitudes for the basis states with~$q_{n-1}$ set to $\ket{0}$; the other entries represent the amplitudes for states with $q_{n-1}$ set to~$\ket{1}$. This decomposition is represented in a decision diagram structure by a node labeled~$q_{n-1}$ and two successors leading to nodes representing the sub-vectors. The sub-vectors are recursively decomposed further until vectors of size~1 (i.e., a complex number) result. This eventually represents the amplitude~$\alpha_i$ for the complete basis state and is given by a terminal node.
During this decomposition, equivalent sub-vectors are represented by the same nodes---enabling sharing and, hence, a reduction of the complexity of the representation.
An example illustrates the idea.

\begin{figure}[tbp]
	\centering
	\begin{subfigure}[b]{0.25\textwidth}
		\centering
		\begin{tikzpicture}[terminal/.style={draw,rectangle,inner sep=0pt}]
		\matrix[ampersand replacement=\&,every node/.style={vertex},column sep={1cm,between origins},row sep={1cm,between origins}] (qmdd) {
			\& \node (n1) {$q_2$}; \& \\
			\node[xshift=0.5cm] (n2a) {$q_1$}; \& \node[xshift=0.5cm] (n2b) {$q_1$}; \& \\
			\node[xshift=0.5cm] (n3a) {$q_0$}; \& \node[xshift=0.5cm] (n3b) {$q_0$}; \& \\[0.2cm]
			\node[terminal] (t0){$0$}; \& \node[terminal] (t1) {$\frac{1}{2}$}; \& \node[terminal] (t2) {$-\frac{1}{\sqrt{2}}$};\\
		};
		
		\draw (n1) -- ++(240:0.6cm) -- (n2a);
		\draw[thick] (n1) -- ++(300:0.6cm) -- (n2b);
		
		\draw (n2a) -- ++(240:0.6cm) -- (t0);
		\draw (n2a) -- ++(300:0.6cm) -- (n3a);

		\draw (n2b) -- ++(240:0.6cm) -- (n3a);
		\draw[thick] (n2b) -- ++(300:0.6cm) -- (n3b);
		
		\draw (n3a) -- ++(240:0.6cm) -- (t1);
		\draw (n3a) -- ++(300:0.6cm) -- (t0);
		
		\draw[thick] (n3b) -- ++(240:0.6cm) -- (t2);
		\draw (n3b) -- ++(300:0.6cm) -- (t0);	
		\end{tikzpicture}
		\caption{Without weights}
		\label{fig:dd_vector_a}		
	\end{subfigure}%
	\begin{subfigure}[b]{0.25\textwidth}
		\centering
		\begin{tikzpicture}[terminal/.style={draw,rectangle,inner sep=0pt}]	
			\matrix[ampersand replacement=\&,every node/.style={vertex},column sep={1cm,between origins},row sep={1cm,between origins}] (qmdd2) {
				\& \node (m1) {$q_2$}; \& \\
				\node[xshift=0.5cm] (m2a) {$q_1$}; \& \node[xshift=0.5cm] (m2b) {$q_1$}; \& \\
				\& \node (m3) {$q_0$}; \& \\
				\& \node[terminal] (t3) {1}; \& \\
			};
			
			\draw[thick] ($(m1)+(0,0.7cm)$) -- (m1) node[right, midway]{$\frac{1}{2}$};
			
			\draw (m1) -- ++(240:0.6cm) -- (m2a);
			\draw[thick] (m1) -- ++(300:0.6cm) -- (m2b);
			
			\draw (m2a) -- ++(240:0.5cm) node[zeroterm] {$0$};
			\draw (m2a) -- ++(300:0.6cm) -- (m3);
	
			\draw (m2b) -- ++(240:0.6cm) -- (m3);
			\draw[thick] (m2b) -- ++(300:0.6cm) -- (m3) node[right, midway] {\bf $-\sqrt{2}$};
			
			\draw[thick] (m3) -- ++(240:0.6cm) -- (t3);
			\draw (m3) -- ++(300:0.5cm) node[zeroterm] {$0$};
		\end{tikzpicture}
		\caption{With weights}
		\label{fig:dd_vector_b}		
	\end{subfigure}%
	\begin{subfigure}[b]{0.5\textwidth}
		\centering
		\begin{tikzpicture}[
				node distance=1 and 0.5,
			    define color/.code={
			        \definecolor{hsb#1}{Hsb}{#1, 1, 0.75}
			    },				
			    edge/.style 2 args={
   					line width={#1pt},
   					define color={#2},
   					draw=hsb#2
   				},
				edge0/.style 2 args={
					line width={#1pt},
					define color={#2},
					draw=hsb#2,
					out=-130, 
					in=90
				},
				edge1/.style 2 args={
					line width={#1pt},
					define color={#2},
					draw=hsb#2,
					out=-50, 
					in=90
				},
				zerostub/.style={
					inner sep=0, 
					minimum size=3pt, 
					circle, 
					fill=black
				},
				on grid
			]
			\node[vertex] (q2) {2};
			\node[vertex, below left=of q2] (q1l) {1};
			\node[vertex, below right=of q2] (q1r) {1};
			\node[vertex, below right=of q1l] (q0) {0};
			\node[terminal, below=of q0] (t) {};
			
			\draw[edge={.5}{0}] ($(q2)+(0,0.7cm)$) -- (q2);
			\draw[edge0={1}{0}] (q2) to (q1l);
			\draw[edge1={1}{0}] (q2) to (q1r);
			\draw[edge1={1}{0}] (q1l) to (q0);
			\draw[black, thick] (q1l) to ++(-130:0.35) node[zerostub]{};
			\draw[edge0={1}{0}] (q1r) to (q0);
			\draw[edge1={1.41}{180}] (q1r) to (q0);
			\draw[edge0={1}{0}] (q0) to (t);
			\draw[black, thick] (q0) -- ++(-50:0.35) node[zerostub]{};
			\begin{scope}[xshift=3cm, yshift=-1cm,
						font=\large,
					   define color/.code={
					       \definecolor{hsb#1}{Hsb}{#1, 1, 0.75}
					   },
					   wheel color/.style={
					   		line width=4mm,
					   		define color={#1},
					   		draw=hsb#1
					   }]
					\foreach \x in {0,2,...,358} {
						\draw [wheel color=\x] (\x:0.9) arc (\x:\the\numexpr\x+3:0.9);
					}
					\node at (0:1.4) {\(0\)};
					\node at (90:1.4) {\(\frac{\pi}{2}\)};
					\node at (180:1.4) {\(\pi\)};
					\node at (270:1.5) {\(\frac{3\pi}{2}\)};
			\end{scope}
		\end{tikzpicture}
		\caption{Graphical notation}
		\label{fig:dd_vector_c}		
	\end{subfigure}%
	\caption{Different representations of the state vector from \autoref{ex:dd_vector}}
	\label{fig:dd_vector}
	\vspace{-1em}
\end{figure}
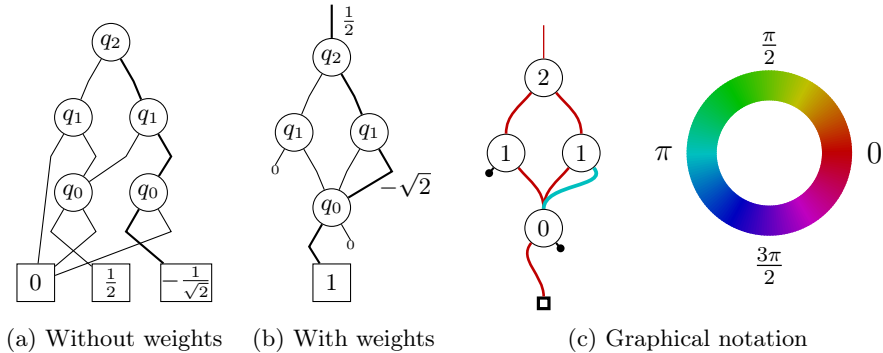

\begin{example}
	\label{ex:dd}
	Consider again the quantum state from \autoref{ex:dd_vector}.
	Applying the decomposition described above yields a decision diagram as shown in \autoref{fig:dd_vector_a}.	
	The left (right) outgoing edge of each node labeled~$q_i$ points to a node representing the sub-vector with all amplitudes for the basis states with~$q_i$ set to $\ket{0}$ ($\ket{1}$).
	Following a path from the root to the terminal yields the respective entry. 
	For example, following the path highlighted bold in \autoref{fig:dd_vector_a} provides the amplitude for the basis state with $q_2=\ket{1}$ (right edge), $q_1=\ket{1}$ (right edge), and $q_0=\ket{0}$ (left edge), i.e., $-\frac{1}{\sqrt{2}}$ which is exactly the amplitude for basis state $\ket{110}$ (seventh entry in the vector from \autoref{ex:dd_vector}).
	Since some sub-vectors are equal (e.g., $\left[\frac{1}{2},0\right]^T$ represented by the left node labeled~$q_0$), sharing is possible. 	
\end{example}

However, there is more potential for sharing. 
In fact, many entries of the state vectors differ by a common factor only (e.g., the state vector from \autoref{ex:dd_vector} has entries~$\frac{1}{2}$ and~$-\frac{1}{\sqrt{2}}$ which differ by the factor~$-\sqrt{2}$).
This is exploited in the decision diagram representation by denoting common factors of amplitudes as weights to the edges of the decision diagram. 
Then, the value of an amplitude for a basis state is determined by not only following the path from the root to the terminal, but
additionally multiplying the weights of the edges along this path. 
Note that for a more readable notation, we use zero stubs to indicate zero vectors (i.e., vectors only containing zeroes) and omit edge weights that are equal to one.
Again, an example illustrates the idea.

\begin{example}
	Consider again the quantum state from \autoref{ex:dd_vector} and the corresponding decision diagram shown in \autoref{fig:dd_vector_a}.	
	As can be seen, the sub-graphs rooting the node labeled~$q_0$ are structurally equivalent and only differ in their terminal values. Moreover, they
	represent sub-vectors~$\left[\frac{1}{2},0\right]^T$ and~$\left[-\frac{1}{\sqrt{2}},0\right]^T$
	which only differ in a common factor.
	
	In the decision diagram shown in \autoref{fig:dd_vector_b}, both sub-graphs are merged. This is possible since the corresponding value of the amplitudes is now determined not by the terminals, but the product of weights on the respective paths. As an example, consider again the path highlighted bold representing the	
	amplitude for the basis state~$\ket{110}$. Since this path includes the weights $\frac{1}{2}$, 1, $-\sqrt{2}$, and $1$,
	an amplitude value of $\frac{1}{2}\cdot 1 \cdot (-\sqrt{2}) \cdot 1 = -\frac{1}{\sqrt{2}}$ results.
\end{example}

There exist various possibilities to factorize an amplitude. 
Hence, we apply a normalization scheme to the decision diagrams, resulting in a representation which is canonical w.r.t~order of qubits~\cite{millerQMDDDecisionDiagram2006}.
The outgoing edges of a node are often normalized by dividing both weights by the weight of the left-most edge (when $\neq 0$), and multiplying this factor
to the incoming edges.
However, it has been found in~\cite{hillmich2020just}, that it is more effective to divide by the norm of the vector containing both edge weights and adjust the incoming edges accordingly.
This normalizes the sum of the squared magnitudes of the outgoing edge weights to~1
and is consistent with the quantum semantics, where basis states \(\ket{0}\) and \(\ket{1}\) are observed after measurement with probabilities that are squared magnitudes of the respective weights.
Furthermore, to ease the graphical notation we represent the complex number in polar plane as \(r \cdot \mathrm{e}^{\mathrm{i}\alpha}\).
The magnitude \(r\) of an edge weight is represented by the edge's thickness and the angle \(\alpha\) according to the HLS color wheel~\cite{willeVisualizingDecisionDiagrams2021}.
The graphical notation reflects that one is most often only interested in the structure of the decision diagram instead of the exact values of edge weights.
Of course, the edge weights can be put in the notation if necessary.

\begin{example}\label{ex:colorcode}
	Consider again the quantum state from \autoref{ex:dd_vector} and the normalized decision diagram with edge weights shown in \autoref{fig:dd_vector_b}.
	\autoref{fig:dd_vector_c} shows the graphical notation of this decision diagram where the line width represents the magnitude of the edge weight and the color the respecting angle when considering the polar notation of complex numbers.
	
	In \autoref{fig:dd_vector_c}, the edge to the root node (having a weight of \(\nicefrac{1}{2}\)) is notably thinner than the other edges (with weights \(1\) and \(-\sqrt{2}\)).
	The with weight \(-\sqrt{2}\) is slightly thicker than the edges with weight \(1\) and, more visible in the figure, has a different phase, i.e.,~\(-\sqrt{2} = \sqrt{2}\cdot e^{i\pi}\), encoded in the line color.
\end{example}

Overall, the discussions from above lead to the following definition of decision diagrams for quantum states.

\begin{definition}
	\label{def:dd_sv}
	The decision diagram representing a $2^n$-dimensional state vector is a directed acyclic multi-graph with one terminal node labeled 1 that has no successors and represents a 1-dimensional vector with the element 1. 
	All other nodes are labeled $q_i$, $0\le i < n$ (representing a partition over qubit $q_i$) and have two successors. Additionally, there is an edge pointing to the root node of the decision diagram. 
	This edge is called \emph{root edge}. 
	Each edge of the graph has  a complex number attached as weight. 
	An entry of the state vector is then determined by the product of all edge weights along the path from the root towards the terminal. 
	Without loss of generality, the nodes of the decision diagram are ordered by the significance of their label, i.e., the successor of a node labeled $q_i$ are labeled with a less significant qubit $q_j$.
	Finally, the nodes are normalized, which means that the sum of the squared magnitudes of the outgoing edge weights equals one and the common factor is propagated upwards in the decision diagram.
\end{definition}

\subsection{Representation of Matrices}
\label{sec:matrices}

While quantum states are commonly represented by vectors, quantum operations are described by matrices.
These matrices are unitary (its conjugate transpose is also its inverse) and \(2^n\times2^n\)-dimensional for a \(n\)-qubit system.
Similar to state vectors, matrices often include redundancies, which can be exploited for a more compact representation. 
To this end, the decomposition scheme for state vectors is extended by a second dimension -- yielding a decomposition scheme for $2^n\times 2^n$ matrices. 

The entries of a unitary matrix $U = [u_{i,j}]$ indicate how much the operation~$U$ affects the mapping from a basis state~$\ket{i}$ to a basis state~$\ket{j}$.
Considering again a quantum system with qubits $q_{n-1}, \ldots, q_1, q_0$, whereby~$q_{n-1}$ represents the most significant qubit,
the matrix~$U$ is decomposed into four sub-matrices with dimension $2^{n-1}\times 2^{n-1}$: 
All entries in the left upper sub-matrix (right lower sub-matrix) provide the values describing the mapping from basis states $\ket{i}$ to $\ket{j}$ with both
assuming $q_0 =\ket{0}$ ($q_0 =\ket{1}$). All entries in the right upper sub-matrix (left lower sub-matrix) provide the values describing the mapping from basis states $\ket{i}$ with $q_0 =\ket{1}$ to $\ket{j}$ with $q_0 =\ket{0}$ ($q_0 =\ket{0}$ to $q_0 =\ket{1}$).

This decomposition is represented in a decision diagram structure by a node labeled~$q_{n-1}$ and four successors leading to nodes representing the sub-matrices. 
The sub-matrices are recursively decomposed further until a $1\times 1$ matrix~(i.e., a complex number) results. This eventually represents the value~$u_{i,j}$ for the corresponding mapping. 
Also during this decomposition, equivalent sub-matrices are represented by the same nodes and weights.
As for decision diagrams representing state vectors, a corresponding normalization scheme is employed. 
To this end, all edges-weights are divided by the leftmost entry with the largest magnitude.
Again, zero stubs are used to indicate zero matrices (i.e., matrices that contain zeros only) and edge weights equal to one are omitted.
Similar to decision diagrams for quantum states, magnitude and phase of edge weights are encoded as thickness and color, respectively (see \autoref{ex:colorcode}). 
Again, an example illustrates the idea.

\begin{figure}[tbp]
	\begin{subfigure}[b]{0.3\textwidth}
		\centering
		\begin{tikzpicture}[
			define color/.code={
			    \definecolor{hsb#1}{Hsb}{#1, 1, 0.75}
			},				
			edge/.style n args={3}{
   					line width={#1pt},
   					define color={#2},
   					draw=hsb#2,
					out=#3, 
					in=90
  			},
			on grid
		]
		\matrix[matrix of nodes,ampersand replacement=\&,column sep={1cm,between origins},row sep={1cm,between origins}] (qmdd) {
			\node[draw = none] (top) {}; \\
			\node[vertex] (n1) {0}; \\
			\node[terminal] (t){};\\
		};
		
		\draw[edge={1}{0}{-130}] (n1) to (t);
		\draw[edge={1}{0}{-100}] (n1) to (t);
		\draw[edge={1}{0}{-80}] (n1) to (t);
		\draw[edge={1}{180}{-50}] (n1) to (t);
		
		\draw[edge={0.7071}{0}{-90}] (top) -- (n1);
		\end{tikzpicture}
		\caption{$H$}
		\label{fig:dd_h}		
	\end{subfigure}\hfill
	\begin{subfigure}[b]{0.3\textwidth}
		\centering
		\begin{tikzpicture}[
			zerostub/.style={
				inner sep=0, 
				minimum size=3pt, 
				circle, 
				fill=black
			},
			define color/.code={
				\definecolor{hsb#1}{Hsb}{#1, 1, 0.75}
			},	
			edge/.style n args={3}{
  					line width={#1pt},
  					define color={#2},
  					draw=hsb#2,
				out=#3, 
				in=90
 			},
			on grid
		]
		
		\matrix[matrix of nodes,ampersand replacement=\&,column sep={1cm,between origins},row sep={1cm,between origins}] (qmdd) {
			\node[draw = none] (top) {}; \\
			\node[vertex] (n1) {0}; \\
			\node[terminal] (t){};\\
		};
		
		\draw[edge={1}{0}{-130}] (n1) to (t);
		\draw[black, thick] (n1) to ++(-100:0.35) node[zerostub]{};
		\draw[black, thick] (n1) to ++(-80:0.35) node[zerostub]{};
		\draw[edge={1}{0}{-50}] (n1) to (t);
		
		\draw[edge={1}{0}{90}] (top) -- (n1);
		\end{tikzpicture}
		\caption{$\mathbb{I}_2$}
		\label{fig:dd_i}		
	\end{subfigure}\hfill
	\begin{subfigure}[b]{0.3\textwidth}
		\centering
		\begin{tikzpicture}[
					zerostub/.style={
						inner sep=0, 
						minimum size=3pt, 
						circle, 
						fill=black
					},
					define color/.code={
						\definecolor{hsb#1}{Hsb}{#1, 1, 0.75}
					},	
					edge/.style n args={3}{
		  					line width={#1pt},
		  					define color={#2},
		  					draw=hsb#2,
						out=#3, 
						in=90
		 			},
					on grid
				]
		\matrix[matrix of nodes,ampersand replacement=\&,column sep={1cm,between origins},row sep={1cm,between origins}] (qmdd) {
			\node[draw = none] (top) {}; \\
			\node[vertex] (n1) {1}; \\
			\node[vertex] (n2) {0}; \\
			\node[terminal] (t){};\\
		};
		
		\draw[edge={1}{0}{-130}] (n1) to (n2);
		\draw[edge={1}{0}{-100}] (n1) to (n2);
		\draw[edge={1}{0}{-80}, dotted] (n1) to (n2);
		\draw[edge={1}{180}{-50}] (n1) to (n2);

		\draw[edge={1}{0}{-130}, dotted] (n2) to (t);
		\draw[black, thick] (n2) to ++(-100:0.35) node[zerostub]{};
		\draw[black, thick] (n2) to ++(-80:0.35) node[zerostub]{};
		\draw[edge={1}{0}{-50}] (n2) to (t);
		
		\draw[edge={0.7071}{0}{90}] (top) -- (n1);
		\end{tikzpicture}
		\caption{$U=H\otimes I_2$}
		\label{fig:dd_kronecker}		
	\end{subfigure}
	\caption{Representation of matrices}
	\label{fig:dd_matrices}
	\vspace{-1em}
\end{figure}
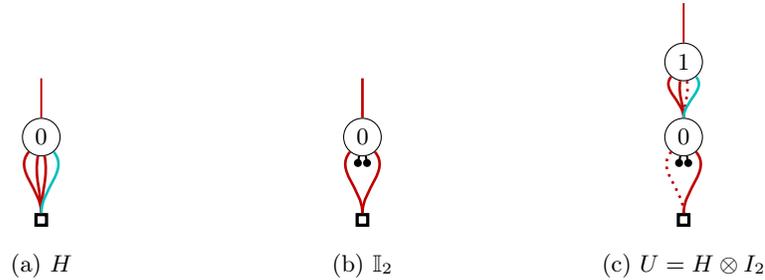

\begin{example}
	\label{ex:kronecker}
	Consider the matrices of the Hadamard operation $H$, the identity~$\mathbb{I}_2$, and their combination $U = H \otimes \mathbb{I}_2$, i.e.,
	\begin{align*}
		H = \frac{1}{\sqrt{2}}\begin{bmatrix*}[r]1 & 1 \\ 1 & -1\end{bmatrix*}
		\quad
		\mathbb{I}_2 = \begin{bmatrix}1 & 0 \\ 0 & 1\end{bmatrix}
		\quad
		U=H\otimes I_2 = \frac{1}{\sqrt{2}}\begin{bsmallmatrix*}[r]
			1 & 0 & 1 & 0 \\
			0 & 1 & 0 & 1 \\
			1 & 0 & -1 & 0 \\
			0 & 1 & 0 & -1 \\
		\end{bsmallmatrix*}.
	\end{align*}
	
	\autoref{fig:dd_matrices} shows the corresponding decision diagram representations.
	Following the path with dotted lines in \autoref{fig:dd_kronecker} defines the entry~$u_{2,0}$: a mapping from $\ket{0}$ to $\ket{1}$ for $q_1$ (third edge from the left) and from $\ket{0}$ to $\ket{0}$ for $q_0$ (first edge). Consequently the path describes the entry for a mapping from $\ket{00}$ to $\ket{10}$.  
	Multiplying all factors on the path (including the \emph{root edge}) yields $\frac{1}{\sqrt{2}}\cdot 1\cdot 1 = \frac{1}{\sqrt{2}}$, which is the value of~$u_{2,0}$.
\end{example}

Overall, the concepts described above yield to the definition of a decision diagram representing a unitary matrix as follows.
\begin{definition}\label{def:dd_matrix}
	The decision diagram representing a $2^n\times 2^n$-dimensional unitary matrix is a directed acyclic graph with one terminal node labeled 1 that has no successors and represents a $1\times 1$ matrix with the element 1. All other nodes are labeled $q_i$, $0\le i < n$ (representing a partition over qubit $q_i$) and have four successors. Additionally, there is an edge pointing to the root node of the decision diagram. This edge is called \emph{root edge}. Each edge of the graph has attached a complex number as weight. An entry of the unitary matrix is then determined by the product of all edge weights along the path from the root towards the terminal. Without loss of generality, the nodes of the decision diagram are ordered by the significance of their label, i.e., the successor of a node labeled $q_i$ are labeled with a less significant qubits $q_j$.
	Finally, the nodes are normalized, which means that all edges-weights are divided by the leftmost entry with the largest magnitude. The common factor is propagated upwards in the decision diagram.
\end{definition}

A performance-oriented implementation handling decision diagrams and operations as described in this section is freely available  under the MIT license at \url{https://github.com/cda-tum/dd_package}.
Furthermore, to give a better intuition and make decision diagrams for quantum computing more accessible, an installation-free web-tool that visualizes decision diagrams for state vectors as well as matrices is available at \url{https://www.cda.cit.tum.de/app/ddvis/}.
\bigskip

Given the decision diagrams as described in this section, the following sections showcase the applicability in different areas of design and verification work.
More precisely, we cover the simulation of quantum circuits without noise in \autoref{sec:sim} and with noise in \autoref{sec:noise} as well as verification of quantum circuits in \autoref{sec:verification}.

\section{Simulation of Quantum Circuits}
\label{sec:sim}

Despite physical quantum computers being available in the cloud nowadays, the simulation of quantum circuits on non-quantum hardware remains paramount for the development and design of future quantum computing applications. 
Additionally, simulations on non-quantum hardware provide insights into the inner workings of a quantum system that are fundamentally hidden in physical quantum computers.
This enables designers to analyze quantum algorithms or verify the output of physical quantum computers.
To this end, simulating a quantum circuit entails the successive application of all individual gates of the circuit to the initial state of a quantum system in order to obtain the final state.
The final state is measured to obtain the result in the computational bases.
While straight-forward in principle, this quickly amounts to a hard task due to the required memory on non-quantum hardware and the subsequently difficult manipulation of $2^n$ complex amplitudes for an $n$-qubit system.

Decision diagrams, as described in \autoref{sec:dds}, provide a promising technique that aims at compactly representing the $2^n$ complex amplitudes of a quantum system and the corresponding operations applied to it.
Having the ability to compactly represent state vectors and unitary matrices, all that is left is to provide corresponding methods to form the Kronecker product, multiply vectors with matrices, as well as measure the quantum system. 
Since the introduced decision diagrams closely relate to vectors and matrices, we can implement the required operations by slight adaptations only.

\subsection{Kronecker Product}\label{sec:kronecker}

The Kronecker product enables composition of multiple matrices to attain the suitable size \mbox{$2^n \times 2^n$} matrix to be applied to an \(n\)-qubit system.
Given two matrices \(A\) and \(B\), the Kronecker product is defined as in \autoref{eqn:kronecker}.
\begin{align}
A\otimes B = \left[ \begin{matrix}
	a_{0,0}\cdot B     & \cdots & a_{0,2^k-1}\cdot B      \\
	\vdots             & \ddots & \vdots                  \\
	a_{2^k-1,0}\cdot B & \cdots & a_{2^k-1,2^k-1} \cdot B \label{eqn:kronecker}
\end{matrix} \right]
\end{align}

In other words, the Kronecker product replaces each element $a_{i,j}$ of $A$ by $a_{i,j}\cdot B$. 
While this constitutes a computationally expensive task using straight-forward realizations by means of array-based implementations of $A$ and $B$, it is very cheap to form the Kronecker product of two matrices given as decision diagrams. 

Since $a_{i,j}$ is given as product of the edge weights from~$A$'s root node to the terminal and we can easily determine $a_{i,j}\cdot B$ by adjusting the weight of the edge pointing to $B$'s root node. All that has to be done to determine $A\otimes B$ is replacing $A$'s terminal with the root node of $B$. Additionally, the weight of $A$'s root edge has to be multiplied by the weight of $B$'s root edge.

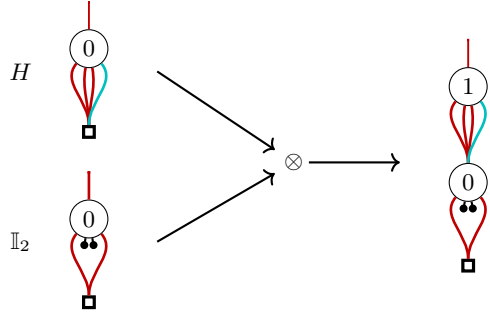
\begin{figure}[tb]
\centering
\resizebox{0.55\linewidth}{!}{
\begin{tikzpicture}[
			zerostub/.style={
				inner sep=0, 
				minimum size=3pt, 
				circle, 
				fill=black
			},
			define color/.code={
				\definecolor{hsb#1}{Hsb}{#1, 1, 0.75}
			},	
			edge/.style n args={3}{
  					line width={#1pt},
  					define color={#2},
  					draw=hsb#2,
				out=#3, 
				in=90
 			}
		]
	\node[draw,circle,inner sep=0pt,minimum width=0.5cm,minimum height=0.5cm] (n1) {$0$};
	\node[below=0.75 of n1, terminal] (t1) {};
	\draw[edge={1}{0}{-130}] (n1) to (t1);
	\draw[edge={1}{0}{-100}] (n1) to (t1);
	\draw[edge={1}{0}{-80}] (n1) to (t1);
	\draw[edge={1}{180}{-50}] (n1) to (t1);
	\draw[edge={0.7071}{0}{90}] ($(n1)+(0,0.6cm)$) to (n1);
				
	\node[draw,circle,inner sep=0pt,minimum width=0.5cm,minimum height=0.5cm, below= 0.8 of t1] (n2) {$0$};
	\node[below=0.75 of n2, terminal] (t2) {};
	\draw[edge={1}{0}{-130}] (n2) to (t2);
	\draw[black, thick] (n2) to ++(-100:0.35) node[zerostub]{};
	\draw[black, thick] (n2) to ++(-80:0.35) node[zerostub]{};
	\draw[edge={1}{0}{-50}] (n2) to (t2);
	\draw[edge={1}{0}{90}] ($(n2)+(0,0.6cm)$) to (n2);
		
	\node[thick] (ot) at ($(t1)!0.35!(n2) + (2.7, 0.0)$) {$\otimes$};
	\draw[->, thick] ($(t1)!0.35!(n2) + (2.9, 0)$) --++(1.2, 0.0);
	\draw[->, thick] ($(n1)+(0.9, -0.3)$) -- (ot);
	\draw[->, thick] ($(n2)+(0.9, -0.3)$) -- (ot);
	
	\node[draw,circle,inner sep=0pt,minimum width=0.5cm,minimum height=0.5cm] (n3) at ($(n1)+(5.0, -0.5)$) {1};
	\node[draw,circle,inner sep=0pt,minimum width=0.5cm,minimum height=0.5cm, below=0.75 of n3] (n4) {0};

	\draw[edge={1}{0}{-130}] (n3) to (n4);
	\draw[edge={1}{0}{-100}] (n3) to (n4);
	\draw[edge={1}{0}{-80}] (n3) to (n4);
	\draw[edge={1}{180}{-50}] (n3) to (n4);
	\draw[edge={0.7071}{0}{90}] ($(n3)+(0,0.6cm)$) to (n3);
				
	\node[below=0.75 of n4, terminal] (t3) {};
	\draw[edge={1}{0}{-130}] (n4) to (t3);
	\draw[black, thick] (n4) to ++(-100:0.35) node[zerostub]{};
	\draw[black, thick] (n4) to ++(-80:0.35) node[zerostub]{};
	\draw[edge={1}{0}{-50}] (n4) to (t3);	
	
	\node (H) at ($(n1)-(0.9, 0.3)$) {$H$};
	\node (I2) at ($(n2)-(0.9, 0.3)$) {$\mathbb{I}_2$};
\end{tikzpicture}}
\caption{Creation of $H\otimes\mathbb{I}_2$ using decision diagrams}
\label{fig:tensordd}
\vspace{-1em}
\end{figure}

\begin{example}
	Recall the matrices considered in \autoref{fig:dd_kronecker}. The Kronecker product $U = H \otimes \mathbb{I}_2$ can efficiently be constructed by taking the decision diagram representation of $H$ and replacing its terminal node with the root node of the decision diagram representing $\mathbb{I}_2$. Since the root edge of $\mathbb{I}_2$ has weight 1, the value of the root node of $U$ is equal to the weight of $A$'s root edge. This is illustrated in \autoref{fig:tensordd}.
\end{example}

\subsection{Adding and Multiplying Unitary Matrices}\label{sec:multiply}

The multiplication of a unitary matrix $U$ and a state vector $\ket{\varphi}$ can be broken down into sub-computations according to \autoref{eqn:multiplication}.
\begin{align}
    \begin{bmatrix}
    	U_{00} & U_{01} \\ U_{10} & U_{11}
    \end{bmatrix}
    \cdot
    \begin{bmatrix}
    	\varphi_{0} \\ \varphi_{1}
    \end{bmatrix} 
    =
    \begin{bmatrix}
    	(U_{00}\varphi_{0} + U_{01}\varphi_{1}) \\ (U_{10}\varphi_{0} + U_{11}\varphi_{1})
    \end{bmatrix}\label{eqn:multiplication}
\end{align}

For decision diagrams, recursively determining the four sub-products $U_{00} \cdot \varphi_0$, $U_{01} \cdot \varphi_1$, $U_{10} \cdot \varphi_0$, and $U_{11} \cdot \varphi_1$ realizes the multiplication.
The decompositions of multiplication and addition are recursively applied until \mbox{$1\times 1$} matrices or 1-dimensional vectors result. 
Since these represent just complex numbers, their multiplication and addition is well defined.

\begin{figure*}[tbp]
	\centering
	\resizebox{0.98\linewidth}{!}{
		\begin{tikzpicture}[terminal/.append style={draw,rectangle,inner sep=2pt}]
			\matrix[ampersand replacement=\&,every node/.style={vertex},column sep={0.8cm,between origins},row sep={1cm,between origins}] (qmdd) {
				\& \& \node (top)[draw = none] {};\& \& \\
				\& \& \node (n1) {$i$};\& \& \\
				\node[dashed, xshift=0.4cm] (n2) {$\phantom{q_i}$}; \& \node[dashed, xshift=0.4cm] (n3) {$\phantom{q_i}$}; \& \node[dashed, xshift=0.4cm] (n4) {$\phantom{q_i}$}; \& \node[dashed, xshift=0.4cm] (n5) {$\phantom{q_i}$}; \& \\
			};
			
			\draw (top) -- (n1);
			
			\draw (n1) -- ++(240:0.6cm) -- (n2);
			\draw (n1) -- ++(260:0.6cm) -- (n3);
			\draw (n1) -- ++(280:0.6cm) -- (n4);
			\draw (n1) -- ++(300:0.6cm) -- (n5);
			
			\matrix[right=0.25cm of qmdd, ampersand replacement=\&,every node/.style={vertex},column sep={0.8cm,between origins},row sep={1cm,between origins}] (qmdd2) {
				\& \node (top2)[draw = none] {};\& \\
				\& \node (m1) {$i$};\& \\
				\node[dashed, xshift=0.4cm] (m2) {$\phantom{q_i}$}; \& \node[dashed, xshift=0.4cm] (m3) {$\phantom{q_i}$}; \& \\
			};
			
			\draw (top2) -- (m1);
			
			\draw (m1) -- ++(240:0.6cm) -- (m2);
			\draw (m1) -- ++(300:0.6cm) -- (m3);				
			
			\matrix[right=0.5cm of qmdd2, ampersand replacement=\&,every node/.style={vertex},column sep={1.5cm,between origins},row sep={1cm,between origins}] (qmdd3) {
				\& \node (top3)[draw = none] {};\& \\
				\& \node (o1) {$i$};\& \\
				\node[dashed, xshift=0.75cm] (o2) {$\phantom{q_i}$}; \& \node[dashed, xshift=0.75cm] (o3) {$\phantom{q_i}$}; \& \\
			};
			
			\draw (top3) -- (o1);
			
			\draw (o1) -- ++(240:0.6cm) -- (o2);
			\draw (o1) -- ++(300:0.6cm) -- (o3);
			
			\matrix[right=-0.25cm of qmdd3, ampersand replacement=\&,every node/.style={vertex},column sep={1.5cm,between origins},row sep={1cm,between origins}] (qmdd4) {
				\& \node (top4)[draw = none] {};\& \\
				\& \node (l1) {$i$};\& \\
				\node[dashed, xshift=0.75cm] (l2) {$\phantom{q_i}$}; \& \node[dashed, xshift=0.75cm] (l3) {$\phantom{q_i}$}; \& \\
			};
			
			\draw (top4) -- (l1);
			
			\draw (l1) -- ++(240:0.6cm) -- (l2);
			\draw (l1) -- ++(300:0.6cm) -- (l3);

			\matrix[right=0.5 of qmdd4, ampersand replacement=\&,every node/.style={vertex},column sep={1.5cm,between origins},row sep={1cm,between origins}] (qmdd5) {
				\& \node (top5)[draw = none] {};\& \\
				\& \node (q1) {$i$};\& \\
				\node[dashed, xshift=0.75cm] (q2) {$\phantom{q_i}$}; \& \node[dashed, xshift=0.75cm] (q3) {$\phantom{q_i}$}; \& \\
			};
			\draw (top5) -- (q1);
			
			\draw (q1) -- ++(240:0.6cm) -- (q2);
			\draw (q1) -- ++(300:0.6cm) -- (q3);

			\draw (n2.south) node[anchor=north] {$U_{00}$};
			\draw (n3.south) node[anchor=north] {$U_{01}$};
			\draw (n4.south) node[anchor=north] {$U_{10}$};
			\draw (n5.south) node[anchor=north] {$U_{11}$};
			
			\draw (m2.south) node[anchor=north] {$\varphi_{0}$};
			\draw (m3.south) node[anchor=north] {$\varphi_{1}$};
			
			\draw (o2.south) node[anchor=north] {$U_{00}\cdot \varphi_{0}$};
			\draw (o3.south) node[anchor=north] {$U_{10}\cdot \varphi_{0}$};
			
			\draw (l2.south) node[anchor=north] {$U_{01}\cdot \varphi_{1}$};
			\draw (l3.south) node[anchor=north] {$U_{11}\cdot \varphi_{1}$};
			
			\draw ($(n5)!0.5!(m2) + (0,0.5)$) node {$\times$};
			
			\draw ($(o3)!0.5!(l2) + (0,0.5)$) node {$+$};
			
			\draw ($(m3)!0.5!(o2) + (0,0.5)$) node {$=$};
			\draw ($(l3)!0.5!(q2) + (0,0.5)$) node {$=$};
			
			\draw ($(q2.south)+(0,-1.2)$) node[anchor=north, label={[align=center]$U_{00}\cdot \varphi_{0}$ \\$+$\\ $U_{01}\cdot \varphi_{1}$}] {};
			\draw ($(q3.south)+(0,-1.2)$) node[anchor=north, label={[align=center]$U_{10}\cdot \varphi_{0}$ \\$+$\\ $U_{11}\cdot \varphi_{1}$}] {};
			
	\end{tikzpicture}}
	\caption{Recursive structure of multiplication and addition using decision diagrams}
	\label{fig:qmdd_mult}
	\vspace{-1em}
\end{figure*}
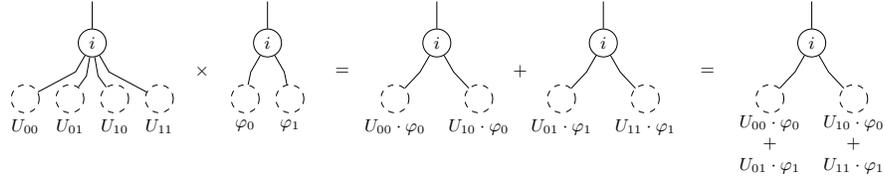

As shown in the middle of \autoref{fig:qmdd_mult}, these sub-products are then combined with a decision diagram node to two intermediate state vectors. 
Finally, these intermediate state vectors have to be added. 
This addition is recursively decomposed similarly, namely as in \autoref{eqn:addition}.
\begin{align}
	\psi + \phi 
	= \begin{bmatrix}\psi_0 \\\psi_1 \end{bmatrix} +  \begin{bmatrix}\phi_0 \\\phi_1 \end{bmatrix} 
	= \begin{bmatrix}\psi_0 + \phi_0 \\ \psi_1 + \phi_1 \end{bmatrix}\label{eqn:addition}
\end{align}
The recursively determined sub-sums $\psi_0 + \phi_0$ and $\psi_1 + \phi_1$ are composed by a decision diagram node as shown on the right-hand side of \autoref{fig:qmdd_mult}.

Moreover, decomposition into sub-products and sub-sums does not change the decision diagram structure. 
Hence, the complexity of them remains bounded by the number of nodes of the original representations. Furthermore, redundancies can again be exploited by caching sub-products and sub-sums. 

\subsection{Measurement}
\label{sec:measure}

Measurement can efficiently be conducted  on the decision diagram structure. 
Without loss of generality, consider that the most significant qubit (which is represented by the root node of the corresponding decision diagram) of the state vector should be measured.
This can be accomplished by applying a SWAP operation or by re-arranging the nodes and edges of the decision diagram.
Then, the probability of choosing either the left or right edge is given by the \emph{upstream} probability of the successor nodes weighted by the corresponding edge weights~\cite{DBLP:journals/tcad/ZulehnerW19}.
Depending on the used normalization scheme, this calculation may be simplified~\cite{DBLP:conf/dac/HillmichMW20}.
An example illustrates the idea.

\begin{example}
	Consider again the quantum state discussed in \autoref{ex:dd_vector} and its corresponding decision diagram shown in \autoref{fig:dd_vector_b}.
	Then, the probabilities for measuring $q_2=\ket{0}$ and $q_2=\ket{1}$ are $|\frac{1}{2}|^2\cdot|1|^2\cdot1 = \frac{1}{4}$ and $|\frac{1}{2}|^2\cdot|1|^2\cdot3 = \frac{3}{4}$, respectively.
\end{example}

Having the probabilities for collapsing the most significant qubit to basis state $\ket{0}$ and $\ket{1}$ allows to sample its new value. 
If we obtain basis state $\ket{0}$ ($\ket{1}$), the amplitudes for all basis states with $q_{n-1}=\ket{1}$ ($q_{n-1}=\ket{0}$) are set to zero. 
In the decision diagram, the collapse is performed by multiplying with the \(M_0\)~(\(M_1\)) non-unitary matrix from \autoref{eqn:measurement-matrices}.
\begin{align}
	M_0 = \begin{bmatrix}1 & 0 \\ 0 & 0\end{bmatrix}
	\quad
	M_1 = \begin{bmatrix}0 & 0 \\ 0 & 1\end{bmatrix}
	\label{eqn:measurement-matrices}
\end{align}
Afterwards the decision diagram is renormalized and the probability of the considered qubit being measurement again in the same basis state in subsequent measurements is 1.

\begin{figure}[tpb]
	\centering
	\begin{subfigure}[b]{0.4\textwidth}
		\centering
		\begin{tikzpicture}[
						node distance=1 and 0.5,
					    define color/.code={
					        \definecolor{hsb#1}{Hsb}{#1, 1, 0.75}
					    },				
					    edge/.style 2 args={
		   					line width={#1pt},
		   					define color={#2},
		   					draw=hsb#2
		   				},
						edge0/.style 2 args={
							line width={#1pt},
							define color={#2},
							draw=hsb#2,
							out=-130, 
							in=90
						},
						edge1/.style 2 args={
							line width={#1pt},
							define color={#2},
							draw=hsb#2,
							out=-50, 
							in=90
						},
						zerostub/.style={
							inner sep=0, 
							minimum size=3pt, 
							circle, 
							fill=black
						},
						on grid
					]
		\matrix[matrix of nodes,ampersand replacement=\&,column sep={1cm,between origins},row sep={1cm,between origins}] (qmdd2) {
			\node[vertex] (m1) {2}; \\
			\node[vertex] (m2b) {1}; \\
			\node[vertex] (m3) {0}; \\
			\node[terminal] (t3) {}; \\
		};
		
		\draw[edge={0.5}{0}] ($(m1)+(0,0.7cm)$) -- (m1) node[right, midway]{$\frac{1}{2}$};
		
		\draw[thick, black] (m1) -- ++(240:0.35) node[zerostub]{};
		\draw[edge1={1}{0}] (m1) to (m2b);
			
		\draw[edge0={1}{0}] (m2b) to (m3);
		\draw[edge1={1.41}{180}] (m2b) to (m3);
		
		\draw[edge0={1}{0}] (m3) to (t3);
		\draw[thick, black] (m3) -- ++(300:0.35) node[zerostub]{};
		
		\end{tikzpicture}
		\caption{Measure $q_2=\ket{1}$}
		\label{fig:dd_measurement_a}		
	\end{subfigure}\hfill
	\begin{subfigure}[b]{0.4\textwidth}
		\centering
		\begin{tikzpicture}[
								node distance=1 and 0.5,
							    define color/.code={
							        \definecolor{hsb#1}{Hsb}{#1, 1, 0.75}
							    },				
							    edge/.style 2 args={
				   					line width={#1pt},
				   					define color={#2},
				   					draw=hsb#2
				   				},
								edge0/.style 2 args={
									line width={#1pt},
									define color={#2},
									draw=hsb#2,
									out=-130, 
									in=90
								},
								edge1/.style 2 args={
									line width={#1pt},
									define color={#2},
									draw=hsb#2,
									out=-50, 
									in=90
								},
								zerostub/.style={
									inner sep=0, 
									minimum size=3pt, 
									circle, 
									fill=black
								},
								on grid
							]
		\matrix[matrix of nodes,ampersand replacement=\&,column sep={1cm,between origins},row sep={1cm,between origins}] (qmdd2) {
			\node[vertex] (m1) {2}; \\
			\node[vertex] (m2b) {1}; \\
			\node[vertex] (m3) {0}; \\
			\node[terminal] (t3) {}; \\
		};
		
		\draw[edge={0.577}{0}] ($(m1)+(0,0.7cm)$) -- (m1)  node[right, midway]{$\frac{1}{\sqrt{3}}$};
		
		\draw[thick, black] (m1) -- ++(240:0.35) node[zerostub]{};
		\draw[edge1={1}{0}] (m1) to (m2b);
			
		\draw[edge0={1}{0}] (m2b) to (m3);
		\draw[edge1={1.41}{180}] (m2b) to (m3);
		
		\draw[edge0={1}{0}] (m3) to (t3);
		\draw[thick, black] (m3) -- ++(300:0.35) node[zerostub]{};
		
		\end{tikzpicture}
		\caption{Normalize amplitudes}
		\label{fig:dd_measurement_b}		
	\end{subfigure}
	\caption{Measurement of qubit $q_2$}
	\label{fig:dd_measurement}
	\vspace{-1.5em}
\end{figure}
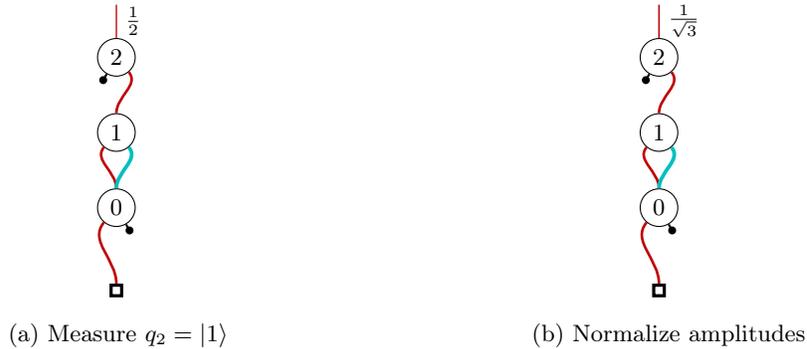

\addtocounter{example}{-1}
\begin{example}[continued]
	Assume we measure basis state~$\ket{1}$ for qubit $q_2$. \autoref{fig:dd_measurement_a} shows the resulting decision diagram. 
	To fulfil the normalization constraint, we renormalize the decision diagram -- eventually resulting in the decision diagram shown in \autoref{fig:dd_measurement_b}.
\end{example}

Measuring all qubits can be conducted in a similar fashion. In fact, we repeat the procedure discussed above sequentially for all qubits $q_{n-1}, q_{n-2}, \ldots, q_{0}$. Assume that qubit $q_i$ shall be measured, and that all qubits $q_{j}$ where $j>i$ are already measured. Then, there exists only one node labeled $q_i$, which is the root node of the sub-vector to be measured. 

Overall, this allow for an more efficient quantum circuit simulation in many cases. An implementation of this method is available
at \url{https://github.com/cda-tum/ddsim} and via the corresponding Python package \href{https://pypi.org/project/mqt.ddsim/}{\texttt{mqt.ddsim}}.
In addition, \url{https://www.cda.cit.tum.de/app/ddvis/} provides an installation-free visualization scheme of the procedure.

\section{Noise-Aware Simulation of Quantum Circuits}\label{sec:noise}

The methods reviewed above allow for an efficient simulation of \emph{perfect}, i.e., error-free, quantum circuits. 
While already an important step towards proper design and evaluation of certain applications, physical quantum computers do not work perfectly and are affected by noise effects, which cause errors during quantum computations.
Considering those errors during simulation enables a more accurate and realistic evaluation of the respectively considered quantum circuits. 
At the same time, considering errors introduces new challenges for the already exponentially hard problem of quantum circuit simulation.
In this section, we review how to conduct noisy simulation with decision diagrams. 
To this end, we first review typical noise effects in quantum computing, i.e., recap frequently occurring errors, and, afterwards, discuss two complementary solutions for noise-aware simulation based on decision diagrams (originally proposed in~\cite{grurl2020considering, grurl2021Stoch}). 

\subsection{Errors in Quantum Computations}

During quantum computations, a large variety of errors may occur and affect the output of the corresponding executions. 
Most prominently, two types of errors are distinguished~\cite{tannuNotAllQubits2019}:
\begin{description}
	\item[{Gate errors}:] Any errors that may alter the originally intended functionality of an operation or may lead to an operation not being executed at all.
	\item[{Decoherence errors}:] Any errors caused by the effect that qubits can only hold information for a limited amount of time.
\end{description}

Gate errors heavily depend on the underlying quantum computer technology and even on the qubits to which the respective operations are applied. 
The effect is that the operation either is not executed at all or that a different operation is employed. Often, they are approximated using depolarization errors~\cite{DBLP:conf/date/KhammassiAFAB17,qiskit} and, hence, defined by altering the qubit to a completely random state~\cite{DBLP:books/daglib/0046438}. 
More detailed descriptions of the respective effects are additionally often provided by the vendors of the respective quantum computer, e.g., in case of IBM at~\cite{ibmErrorRate2019}.

Decoherence Errors occur due to the fragile nature of quantum systems. 
Because of this, qubits can only hold information for a limited time and, hence, qubits in a \mbox{high-energy} state ($\ket{1}$) tend to relax towards a low energy state ($\ket{0}$) (i.e., after a certain amount of time, qubits eventually decay towards~$\ket{0}$). 
Moreover, when a qubit interacts with the environment, further errors (such as phase-flip errors) might occur.

\begin{example}\label{ex:error_effects}
Consider a 2-qubit system which is in state \mbox{$\ket{\psi^\prime} = \tfrac{1}{\sqrt{2}}(\ket{00} + \ket{11})$} 
and assume that a gate error might affect this state with probability~$p$. Then, with probability $1-p$, nothing happens (the state remains unchanged) while, with probability $p$, a certain error effect is imposed. Both scenarios can be captured by either employing an \mbox{I}-operation or an operation describing the error effect (e.g., a polarization using \mbox{X}, \mbox{Y}, or \mbox{Z} or a completely random effect), respectively. 

In a similar fashion, consider the same quantum state but assume a decoherence error at the second qubit (more precisely, a dampening error which makes the second qubit decay to $\ket{0}$) with a probability~$p=0.3$. 
Then, a measurement of this state would not lead to $\ket{00}$ or $\ket{11}$ with equal probabilities anymore (as in the error-free case), but to $\ket{00}$ in 50\% of the cases, $\ket{10}$ in 15\% of the cases, and $\ket{11}$ in 35\% of the cases. That is, the probability that the second qubit decays to $\ket{0}$ is substantially larger due to the decoherence.
\end{example}

Overall, errors effects can be seen as (unwanted) operations employed on a quantum system. Accordingly, they could in principle be simulated like any other quantum operation---using, e.g., the methods described before in \autoref{sec:sim}. The main challenge, however, is that whether an error effect happens or not depends on certain probabilities. These need to be captured during the simulation. Existing solutions doing that have been proposed, e.g., in~\cite{qiskit,atos2016,qxSimulator2017,DBLP:journals/corr/WeckerS14,cirq_developers_2021_4586899,jones2018quest,DBLP:journals/corr/SmelyanskiySA16,villalonga2019flexible,forest}.

\subsection{Simulation Methods Using Decision Diagrams}

In this section, we review how decision diagrams may help in providing a solution for noise-aware quantum circuit simulation. The first solution thereby relies on a stochastic approach which employs the main concepts reviewed in \autoref{sec:sim}, while the second solution aims for a deterministic consideration of noise---requiring a more complex representation of quantum states and operations.

More precisely, the first solution (stochastic simulation as proposed in~\cite{grurl2021Stoch}) is based on the vanilla decision diagram-based circuit simulator described in \autoref{sec:sim}. Then, whenever the considered quantum computer might make an error during its simulated operation, it either mimics the effect of the error by additionally employing an error operation (with corresponding probability~$p$) or leaves the state untouched (with probability~$1-p$). 
Consequently, a \emph{single} output state is sampled from the whole spectrum of possible output states by such a run.
 
By iteratively sampling sufficiently many output states (using, e.g., stochastic Monte-Carlo approximation) a rather accurate approximation of the quantum circuit's behavior under the influence of noise effects can be obtained.
The benefit of this approach is that it does not substantially increase the complexity of individual simulation runs when compared to the error-free circuit simulation. Furthermore, individual simulations are independent and, hence, can be executed in parallel.
However, this approach remains stochastic, i.e., it cannot guarantee the best possible accuracy (although evaluations summarized in~\cite{DBLP:conf/date/GrurlKFW21} show that a sufficient accuracy can be achieved for practically relevant use cases).

If an exact consideration of noise is desired, a more elaborate solution is required which describes all noise effects in a deterministic fashion. To this end, the representation of quantum states and quantum operations in terms of vectors and matrices (as used thus far) is not sufficient anymore. More precisely, a description is needed which incorporates all possible states a quantum system may reside in (including the original state but also states resulting from any noise effects with certain probabilities). 

This is accomplished by extending the state vector representation to \emph{density matrices} (also known as \emph{density operators})~\cite{DBLP:conf/iccad/GrurlFW20}).
More precisely, let $\ket\phi$ be a complex vector representing the state of a quantum system. Then, the corresponding \emph{density matrix} is defined as \(\rho = \ket{\phi}\bra{\phi} \mbox{ with } \bra{\phi} \coloneqq \ket{\phi}^{\dagger}\).

\begin{example}
\label{exp:density_matrix}
Consider again the quantum state \mbox{$\ket{\psi^\prime} = \frac{1}{\sqrt{2}}(\ket{00} + \ket{11})$} from \autoref{ex:error_effects}.
The corresponding \emph{density matrix} $\rho$ is given by 
\begin{align}
	\begin{bmatrix}
	    \frac{1}{\sqrt{2}} \\
	    0 \\
	    0 \\
	    \frac{1}{\sqrt{2}} \\
	\end{bmatrix} 
	\cdot
	\begin{bmatrix}
	    \frac{1}{\sqrt{2}} & 0 & 0 & \frac{1}{\sqrt{2}} \\
	\end{bmatrix} 
	=
	\begin{bmatrix}
	    \cellcolor{gray!40}\frac{1}{2} & 0 & 0 & \frac{1}{2}  \\
	    0 & \cellcolor{gray!40}0 & 0 & 0  \\
	    0 & 0 & \cellcolor{gray!40}0 & 0  \\
	    \frac{1}{2} & 0 & 0 & \cellcolor{gray!40}\frac{1}{2}
	\end{bmatrix}.
\end{align}
This representation properly describes the quantum state while, additionally, allowing to store information about the noise-effects on the state. For example, the diagonal entries encode the probabilities for measuring $\ket{00}, \ket{01}, \ket{10},$ and $\ket{11}$, respectively, which is in line with the probabilities obtained from the state vector representation ($\vert\frac{1}{\sqrt{2}}\vert^2 = \frac{1}{2}$).
\end{example}

Based on this representation, various error effects can now be applied by a tuple $(E_0,E_1, \dots ,E_m)$ of \emph{Kraus matrices} satisfying the condition
\begin{align}
    \sum_{i=0}^m E_i^{\dagger}E_i = \mathbb{I}.\label{eq:kraus_condition}
\end{align}
For example, a decoherence error which makes a qubit decay to $\ket{0}$ can be represented by~\cite{DBLP:books/daglib/0046438}
\begin{gather}
  \begin{aligned}
    (E_0, E_1) \text{ with } E_0 = \begin{bmatrix} 1 & 0\\ 0 & \sqrt{1-p} \end{bmatrix} \text{ and } E_1=\begin{bmatrix} 0 & \sqrt{p}\\ 0 & 0 \end{bmatrix}, \label{eq:T1}
  \end{aligned} 
\end{gather}
where the variable $p$ represents the probability of the error occurring.
Other noise effects can be described in a similar fashion.
Applying these error descriptions to a quantum system given by the density matrix $\rho$ yields the density matrix~\cite{DBLP:books/daglib/0046438}:
\begin{gather}
  \begin{aligned}
    \rho^{\prime} = \sum_{i=0}^m E_i \rho E_i^{\dagger}.\label{eq:apply_error}
  \end{aligned} 
\end{gather}

This formalism allows to deterministically capture all noise effects as illustrated in the following example:
\begin{example}
\label{exp:applying_noise}
Consider again the quantum state from above and the decoherence error from \autoref{ex:error_effects} making the second qubit decay to $\ket{0}$) with probability~\mbox{$p=0.3$}.
Then, the error's effect can be calculated deterministically by constructing the correctly sized matrices with the Kronecker product and summing them as in \autoref{eqn:decoherence}.
\begin{align}
	\footnotesize
	\setlength{\arraycolsep}{4pt}
	\medmuskip = 3mu 
	\underbrace{\begin{bmatrix}
		0.5&0&0&0.418\\
		0&0&0&0\\
		0&0&0&0\\
		0.418&0&0&0.35
	\end{bmatrix}}_{\mathit{E_0 \rho E_0^{\dagger}}} 
	+
	\underbrace{\begin{bmatrix}
		0&0&0&0\\
		0&0&0&0\\
		0&0&0.15&0\\
		0&0&0&0
	\end{bmatrix}}_{\mathit{E_1 \rho E_1^{\dagger}}} 
	=
	\underbrace{\left[\begin{array}{cccc}
		\cellcolor{gray!40}{0.5}&0&0&0.418\\
		0&\cellcolor{gray!40}{0}&0&0\\
		0&0&\cellcolor{gray!40}{0.15}&0\\
		0.418&0&0&\cellcolor{gray!40}{0.35}
	\end{array}\right]}_{\mathit{\rho^{\prime}}}\label{eqn:decoherence}
\end{align}
Again the diagonal encodes the probabilities for measuring $\ket{00}, \ket{01}, \ket{10},$ and $\ket{11}$, which are in-line with the values covered before in \autoref{ex:error_effects}.
\end{example}

The resulting representations are substantially larger than the vectors and matrices needed for error-free quantum circuit simulation. For example, rather than vectors of size $2^n$, matrices of size $2^n \times 2^n$ are needed to represent an \(n\)-qubit quantum state. 
However, density matrices can be represented in terms of decision diagrams as well.
In fact, employing the same decomposition scheme for matrices as reviewed in \autoref{sec:sim} yields corresponding decision diagrams for density matrices.

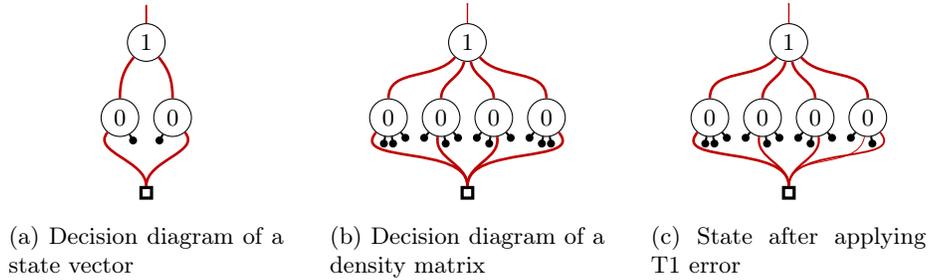
\begin{figure}[t]
	\centering
	\begin{subfigure}[b]{0.3\linewidth}
		\centering
		\begin{tikzpicture}[
					zerostub/.style={
						inner sep=0, 
						minimum size=3pt, 
						circle, 
						fill=black
					},
					define color/.code={
						\definecolor{hsb#1}{Hsb}{#1, 1, 0.75}
					},	
					edge/.style n args={3}{
		  					line width={#1pt},
		  					define color={#2},
		  					draw=hsb#2,
						out=#3, 
						in=90
		 			}
				]
		\matrix[ampersand replacement=\&,column sep={0.35cm,between origins},row sep={1cm,between origins}] (qmdd2) {
			\& \node[vertex] (m1) {1}; \& \\
			\node[vertex] (m2a) {0}; \& \&\node[vertex] (m2b) {0}; \\
			\& \node[terminal] (t) {}; \& \\
		};
		
		\draw[edge={0.7071}{0}{90}] ($(m1)+(0,0.5cm)$) -- (m1);	
		
		\draw[edge={1}{0}{-130}] (m1) to (m2a);
		\draw[edge={1}{0}{-50}] (m1) to (m2b);
		
		\draw[edge={1}{0}{-130}] (m2a) to (t);
		\draw[thick, black] (m2a) -- ++(300:0.35) node[zerostub]{};
		
		\draw[thick, black] (m2b) -- ++(240:0.35) node[zerostub]{};
		\draw[edge={1}{0}{-50}] (m2b) to (t);;
				
		\end{tikzpicture}
		\caption{Decision diagram of a state vector}
		\label{fig:dd_rep_vector}	
	\end{subfigure}\hfill%
	\begin{subfigure}[b]{0.3\linewidth}
		\centering
		\begin{tikzpicture}[
							zerostub/.style={
								inner sep=0, 
								minimum size=3pt, 
								circle, 
								fill=black
							},
							define color/.code={
								\definecolor{hsb#1}{Hsb}{#1, 1, 0.75}
							},	
							edge/.style n args={3}{
				  					line width={#1pt},
				  					define color={#2},
				  					draw=hsb#2,
								out=#3, 
								in=90
				 			}
						]
		\matrix[ampersand replacement=\&,column sep={0.35cm,between origins},row sep={1cm,between origins}] (qmdd2) {
			\&\&\& \node[vertex] (m1) {1}; \&\&\& \\
			\node[vertex] (m2a) {0}; \& \&\node[vertex] (m2b) {0};  \& \& \node[vertex] (m2c) {0}; \& \&\node[vertex] (m2d) {0}; \\
			\&\&\& \node[terminal] (t) {}; \&\&\& \\
		};
		
		\draw[edge={0.5}{0}{90}] ($(m1)+(0,0.5cm)$) to (m1);
		
		\draw[edge={1}{0}{-130}] (m1) to (m2a);
		\draw[edge={1}{0}{-100}] (m1) to (m2b);
		\draw[edge={1}{0}{-80}] (m1) to (m2c);
		\draw[edge={1}{0}{-50}] (m1) to (m2d);
		
		\draw[edge={1}{0}{-130}] (m2a) to (t);
		\draw[thick, black] (m2a) -- ++(-100:0.35) node[zerostub]{};
		\draw[thick, black] (m2a) -- ++(-80:0.35) node[zerostub]{};
		\draw[thick, black] (m2a) -- ++(-50:0.35) node[zerostub]{};
		
		\draw[thick, black] (m2b) -- ++(-130:0.35) node[zerostub]{};
		\draw[edge={1}{0}{-100}] (m2b) to (t);
		\draw[thick, black] (m2b) -- ++(-80:0.35) node[zerostub]{};
		\draw[thick, black] (m2b) -- ++(-50:0.35) node[zerostub]{};
		
		\draw[thick, black] (m2c) -- ++(-130:0.35) node[zerostub]{};
		\draw[thick, black] (m2c) -- ++(-100:0.35) node[zerostub]{};
		\draw[edge={1}{0}{-80}] (m2c) to (t);
		\draw[thick, black] (m2c) -- ++(-50:0.35) node[zerostub]{};
		
		\draw[thick, black] (m2d) -- ++(-130:0.35) node[zerostub]{};
		\draw[thick, black] (m2d) -- ++(-100:0.35) node[zerostub]{};
		\draw[thick, black] (m2d) -- ++(-80:0.35) node[zerostub]{};	
		\draw[edge={1}{0}{-50}] (m2d) to (t);
		\end{tikzpicture}
		\caption{Decision diagram of a density matrix}
		\label{fig:dd_rep_matrix}		
	\end{subfigure}\hfill%
		\begin{subfigure}[b]{0.3\linewidth}
		\centering
		\begin{tikzpicture}[
							zerostub/.style={
								inner sep=0, 
								minimum size=3pt, 
								circle, 
								fill=black
							},
							define color/.code={
								\definecolor{hsb#1}{Hsb}{#1, 1, 0.75}
							},	
							edge/.style n args={3}{
				  					line width={#1pt},
				  					define color={#2},
				  					draw=hsb#2,
								out=#3, 
								in=90
				 			}
						]
		\matrix[ampersand replacement=\&,column sep={0.35cm,between origins},row sep={1cm,between origins}] (qmdd2) {
			\&\&\& \node[vertex] (m1) {1}; \&\&\& \\
			\node[vertex] (m2a) {0}; \& \&\node[vertex] (m2b) {0};  \& \& \node[vertex] (m2c) {0}; \& \&\node[vertex] (m2d) {0}; \\
			\&\&\& \node[terminal] (t) {}; \&\&\& \\
		};
		
		\draw[edge={0.5}{0}{90}] ($(m1)+(0,0.5cm)$) to (m1);
		
		\draw[edge={1}{0}{-130}] (m1) to (m2a);
		\draw[edge={1}{0}{-100}] (m1) to (m2b);
		\draw[edge={1}{0}{-80}] (m1) to (m2c);
		\draw[edge={1}{0}{-50}] (m1) to (m2d);
		
		\draw[edge={1}{0}{-130}] (m2a) to (t);
		\draw[thick, black] (m2a) -- ++(-100:0.35) node[zerostub]{};
		\draw[thick, black] (m2a) -- ++(-80:0.35) node[zerostub]{};
		\draw[thick, black] (m2a) -- ++(-50:0.35) node[zerostub]{};
		
		\draw[thick, black] (m2b) -- ++(-130:0.35) node[zerostub]{};
		\draw[edge={0.836}{0}{-100}] (m2b) to (t);
		\draw[thick, black] (m2b) -- ++(-80:0.35) node[zerostub]{};
		\draw[thick, black] (m2b) -- ++(-50:0.35) node[zerostub]{};
		
		\draw[thick, black] (m2c) -- ++(-130:0.35) node[zerostub]{};
		\draw[thick, black] (m2c) -- ++(-100:0.35) node[zerostub]{};
		\draw[edge={0.836}{0}{-80}] (m2c) to (t);
		\draw[thick, black] (m2c) -- ++(-50:0.35) node[zerostub]{};
		
		\draw[thick, black] (m2d) -- ++(-130:0.35) node[zerostub]{};
		\draw[edge={0.3}{0}{-100}] (m2d) to (t);
		\draw[thick, black] (m2d) -- ++(-80:0.35) node[zerostub]{};	
		\draw[edge={0.7}{0}{-50}] (m2d) to (t);
		\end{tikzpicture}
		\caption{State after applying T1 error}
		\label{fig:dd_after_to_noise}
	\end{subfigure}	
	\caption{Decision diagram representation of states}
	\label{fig:dd_rep_examples}
	\vspace{-1em}
\end{figure}

\begin{example}
	Consider again the quantum state from \autoref{exp:density_matrix} in both the vector as well as density matrix representation. 
	The corresponding decision diagram representations are provided in \autoref{fig:dd_rep_vector} and \autoref{fig:dd_rep_matrix}, respectively.
	The decision diagram resulting after applying the error affect (as considered in \autoref{exp:applying_noise}) is shown in \autoref{fig:dd_after_to_noise}.
\end{example}

Obviously, the decision diagram representations of the density matrices (providing a deterministic representation of all employed error effects) is larger than the original state representation. After all, a substantially larger amount of information needs to be stored. Nevertheless, the examples show that, also in these cases, decision diagrams may offer a more compact representation than offered by a direct representation in terms of a $2^n \times 2^n$-matrix. After all, this helps in improving deterministic, noise-aware quantum circuit simulation.

Overall, this section showed that decision diagrams can be employed for noise-aware quantum circuit simulation---both, stochastically as well as deterministically.
Further details and evaluations on the respective methods are available in~\cite{grurl2021Stoch, grurl2020considering}.
An implementation of the stochastic method is available at \url{https://github.com/cda-tum/ddsim}.

\section{Verification of Quantum Circuits}\label{sec:verification}

As a final example for the utilization of decision diagrams in design/software for quantum computing, we consider the task of verification -- more precisely equivalence checking.
Here, the question is whether two quantum circuits $G$ and $G'$ realize the same functionality.
This is motivated by the design flow in which a given circuit is decomposed, mapped, and optimized~\cite{barencoElementaryGatesQuantum1995, maslovAdvantagesUsingRelative2016, willeImprovingMappingReversible2013, muraliNoiseadaptiveCompilerMappings2019, siraichiQubitAllocation2018,zulehnerEfficientMethodologyMapping2019, willeMappingQuantumCircuits2019, liTacklingQubitMapping2019, matsuoReducingOverheadMapping2019, itokoOptimizationQuantumCircuit2020, vidalUniversalQuantumCircuit2004}.
During all these steps, it has to be ensured that the functionality of the correspondingly resulting circuit descriptions does not change.
In the following, we first give an explicit description of this problem and, then, describe two complementary approaches for tackling it using decision diagrams. %

\subsection{The Quantum Circuit Equivalence Checking Problem}\label{sec:ecproblem}

Equivalence checking in the domain of quantum computing---as we consider it in this work---is about proving that two quantum circuits~$G$ and~$G^\prime$ are functionally equivalent (i.e., realize the same function), or to show the non-equivalence of these circuits by means of a counterexample. 
To this end, consider two quantum circuits \mbox{$G=g_0\dots g_{m-1}$} and \mbox{$G^\prime=g^\prime_0\dots g^\prime_{m^\prime-1}$} operating on $n$ qubits. 
Then, the functionality of each circuit can be uniquely described by the respective system matrices \mbox{$U=U_{m-1}\cdots U_0$} and \mbox{$U^\prime=U_{m^\prime-1}^\prime\cdots U^\prime_0$}, where the matrices~$U^{(\prime)}_i$ describe the functionality of the \mbox{$i$-th}~gate of the respective circuit (with $ 0 \le i < m^{(\prime)}$). %
Consequently, deciding the equivalence of both computations amounts to comparing the system matrices $U$ and $U^\prime$. More precisely, $U$ and $U^\prime$ are considered equivalent, if they at most differ by a global phase factor (which is fundamentally unobservable~\cite{nielsenQuantumComputationQuantum2010}), i.e.,~\mbox{$U = e^{i\alpha}U^\prime$} with $\alpha\in[0,2\pi)$.
\begin{figure}[t]
	\centering
	\begin{subfigure}[b]{0.45\linewidth}
		\centering
		\resizebox{0.9\linewidth}{!}{
			\begin{quantikz}[column sep=5pt, row sep={0.65cm,between origins}, ampersand replacement=\&]
				\lstick{$q_2$} \& \gate{H} \& \phase[label position = above]{\frac{\pi}{2}} \& \phase[label position = above]{\frac{\pi}{4}} \& \qw \& \qw \& \qw \& \swap{2} \& \qw \\
				\lstick{$q_1$} \& \qw \& \ctrl{-1} \& \qw \& \gate{H} \& \phase[label position = above]{\frac{\pi}{2}} \& \qw \& \qw \& \qw\\
				\lstick{$q_0$} \& \qw \& \qw \&  \ctrl{-2} \& \qw \& \ctrl{-1} \& \gate{H} \& \targX{}\& \qw
			\end{quantikz}
		}
		\caption{QFT Circuit $G$}
		\label{fig:qftcirc}
	\end{subfigure}%
	\quad
	\begin{subfigure}[b]{0.45\linewidth}
		\centering
		\resizebox{0.75\linewidth}{!}{
			$\frac{1}{\sqrt{8}}
			\begin{bmatrix}
				1 & 1 & 1 & 1 & 1 & 1 & 1 & 1 \\
				1 & \omega & \omega^2 & \omega^3 & \omega^4 & \omega^5 & \omega^6 & \omega^7\\
				1 & \omega^2 & \omega^4 & \omega^6 & 1 & \omega^2 & \omega^4 & \omega^6\\
				1 & \omega^3 & \omega^6 & \omega^1 & \omega^4 & \omega^7 & \omega^2 & \omega^5\\
				1 & \omega^4 & 1 & \omega^3 & 1 & \omega^4 & 1 & \omega^4\\
				1 & \omega^5 & \omega^2 & \omega^7 & \omega^4 & \omega & \omega^6 & \omega^3\\
				1 & \omega^6 & \omega^4 & \omega^2 & 1 & \omega^6 & \omega^4 & \omega^2\\
				1 & \omega^7 & \omega^6 & \omega^5 & \omega^4 & \omega^3 & \omega^2 & \omega
			\end{bmatrix}$
		}
		\caption{Functionality $U$ ($\omega = (1+i)/\sqrt{2}$)}
		\label{fig:qftmatrix}
	\end{subfigure}%
	
	\begin{subfigure}[b]{\linewidth}
		\centering
		\resizebox{\linewidth}{!}{
			\begin{quantikz}[column sep=4pt, row sep={0.65cm,between origins}, ampersand replacement=\&]
				\lstick{$q_2$} \& \gate{H}\slice{} \& \phase[label position = above]{\frac{\pi}{4}} \& \ctrl{1} \& \qw \& \ctrl{1} \& \qw\slice{} \& \phase[label position = above]{\frac{\pi}{8}} \& \ctrl{2} \& \qw \& \ctrl{2} \& \qw \slice{}\& \qw\slice{}\& \qw\& \qw\& \qw\& \qw\& \qw\slice{}\& \qw\slice{} \& \ctrl{2} \& \targ{} \& \ctrl{2} \& \qw \\
				\lstick{$q_1$} \& \qw \& \qw \& \targ{} \& \phase[label position = above]{-\frac{\pi}{4}} \& \targ{} \& \phase[label position = above]{\frac{\pi}{4}} \& \qw\& \qw\& \qw\& \qw\& \qw\& \gate{H} \& \phase[label position = above]{\frac{\pi}{4}} \& \ctrl{1} \& \qw \& \ctrl{1} \& \qw\& \qw\& \qw\& \qw\& \qw\& \qw \\
				\lstick{$q_0$} \& \qw\&\qw\&\qw\&\qw\&\qw\&\qw\&\qw\&\targ{}\& \phase[label position = above]{-\frac{\pi}{8}} \& \targ{} \& \phase[label position = above]{\frac{\pi}{8}} \& \qw \& \qw \& \targ{} \& \phase[label position = above]{-\frac{\pi}{4}} \& \targ{} \& \phase[label position = above]{\frac{\pi}{4}} \& \gate{H} \& \targ{} \& \ctrl{-2} \& \targ{} \& \qw
			\end{quantikz}
		}
		\caption{Alternative realization $G'$}
		\label{fig:qftdecomp}
	\end{subfigure}%
	\hfill
	\caption{The QFT, its functionality, and an alternative realization}
	\label{fig:qft}
	\vspace{-1em}
\end{figure}
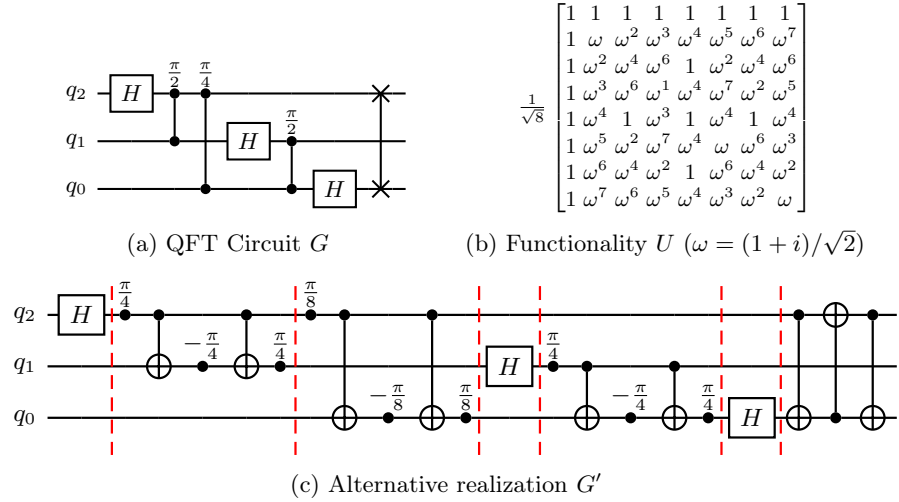

\begin{example}\label{ex:eq}
Consider the circuit $G$ of the three qubit Quantum Fourier Transform shown in \autoref{fig:qftcirc}.
	Its corresponding functionality is described by the densely-populated $8\times 8$ matrix $U$ shown in \autoref{fig:qftmatrix}~\cite{nielsenQuantumComputationQuantum2010}.
	Additionally, \autoref{fig:qftdecomp} shows an alternative realization $G'$ of the functionality of $G$.
	Since both circuits exhibit the same system matrix $U$, they are considered equivalent.
\end{example}

Unfortunately, the whole functionality $U$ (and similarly~$U^\prime$) is not readily available for performing this comparison, but has to be constructed from the individual gate descriptions~$g_i$---requiring the subsequent  matrix-matrix multiplications 
\(U^{(0)} = U_0,\quad U^{(j)} = U_{j} \cdot U^{(j-1)} \mbox{ for } j=1,\dots,m-1\)
to construct the whole system matrix $U = U^{(m-1)}$.
While conceptually simple (as the matrix-vector multiplication for simulation discussed in \autoref{sec:sim}), this quickly constitutes an extremely complex task due to the exponential size of the involved matrices with respect to the number of qubits.
 In fact, equivalence checking of quantum circuits has been shown to be QMA-complete~\cite{janzingNonidentityCheckQMAcomplete2005}.

Due to their potential for compactly representing and efficiently manipulating the functionality of a quantum circuit, decision diagrams are a perfect fit for this task.
However, merely using decision diagrams to construct a representation of both circuits' functionality and comparing them, still has significant shortcomings.
In fact, representing the entire functionality of a quantum circuit still might be exponential in the worst case. 
However, this can be addressed by additionally exploiting the reversibility of quantum circuits.

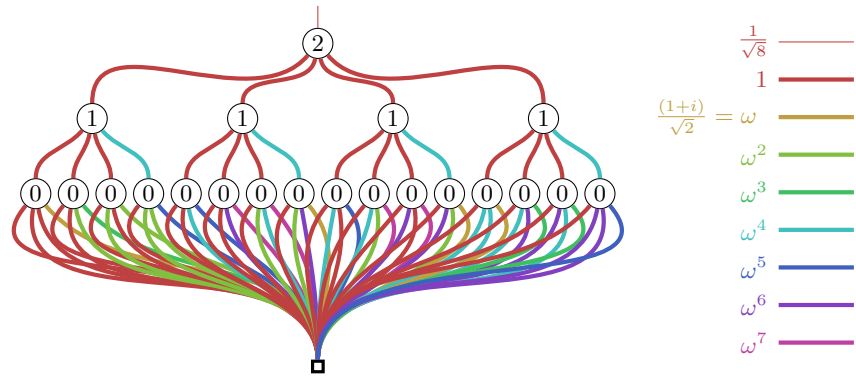
\begin{figure}[t]
\centering
\begin{tikzpicture}
\node (dd) {\usebox{\qftdd}};
\node[right=0.1 of dd] (leg) {\usebox{\legend}};
\end{tikzpicture}
\caption{Decision diagram for the functionality $U$ of $G$ shown in \autoref{fig:qft}}
\label{fig:qftdd}
\vspace{-1em}
\end{figure}

\begin{example}\label{ex:qftdd}
Consider the functionality $U$ of the QFT circuit $G$ from \autoref{fig:qft}. Its corresponding decision diagram (shown in \autoref{fig:qftdd} with color legend to the right-hand side) is as densely populated as the matrix it represents since no redundancy can be exploited, i.e., each node's successors point to unique child nodes.
\end{example}

\subsection{Exploiting Reversibility}\label{sec:exploitreversible}

Most classical logic operations are not reversible (e.g., neither \mbox{$x \land y = 0$} nor \mbox{$x \lor y = 1$} allows one to determine the values of $x$ and $y$).
As there is no bijective mapping between input and output states, in general, the concept of the \emph{inverse} of a classical operation (or a sequence thereof) is not meaningful.
In contrast, all quantum operations are inherently \emph{reversible}. 
Consider an operation $g$ described by the unitary matrix $U$.
Then, its inverse $U^{-1}$ is efficiently calculated as the conjugate-transpose $U^\dag$.
Given a sequence of $m$ operations $g_0, \dots, g_{m-1}$ with associated matrices $U_0,\dots, U_{m-1}$,
the inverse of the corresponding system matrix $U=U_{m-1}\cdots U_0$ is derived by reversing the operations' order and inverting each individual operation, i.e., 
$U^{-1} =U^\dag= U_{0}^\dag \cdots U_{m-1}^\dag$. 

This characteristics can be exploited to improve the performance of the verification approach presented above. 
To this end, consider two quantum circuits $G$ and $G^\prime$. In case both circuits are functionally equivalent, this allows for the conclusion that concatenating one circuit with the inverse of the other realizes the identity function $\mathbb{I}$, i.e., ${G^{\prime -1} \cdot G} \equiv \mathbb{I}$.
This offers significant potential since the identity constitutes the best case for decision diagrams (the identity can be represented by a decision diagram of linear size).
Unfortunately, creating such a concatenation in a naive fashion, e.g., by computing $U \cdot U^{\prime\dag}$ hardly yields any advantage because, even if the final decision diagram would be as compact as possible, the full (and potentially exponential) decision diagram of at least one of the circuits would be generated as an intermediate result.

Instead, the full potential of this observation is utilized if the associativity of the respective multiplications is fully exploited.
More precisely, given two quantum circuits $G$ and $G^\prime$, it holds that
\begin{align*}
G^{\prime -1} \cdot G &=  {(g^{\prime -1}_{m^\prime-1}\dots g^{\prime -1}_0) \cdot (g_0\dots g_{m-1})} \\
&\equiv {(U_{m-1}\cdots U_0)\cdot(U_0^{\prime \dag} \cdots U_{m^\prime -1}^{\prime \dag})} \\
& = {U_{m-1}\cdots U_0\cdot \mathbb{I} \cdot U_0^{\prime \dag} \cdots U_{m^\prime-1}^{\prime \dag}} \\
&\eqqcolon G \shortrightarrow \mathbb{I} \shortleftarrow G^{\prime}.\vspace*{-3mm}
\end{align*}
Here, $G \shortrightarrow \mathbb{I} \shortleftarrow G^{\prime}$ symbolizes that, starting from the identity~$\mathbb{I}$, either gates from $G$ can be ``applied from the left'', or (inverted) gates from $G^{\prime}$ can be ``applied from the right''.
If the respective gates of $G$ and $G^{\prime}$ are applied in a fashion frequently yielding the identity, the entire equivalence checking process can be conducted on rather small (intermediate) decision diagrams. This is illustrated by the following example. 

\begin{example} \label{ex:proposed_approach}
	Consider again the two circuits $G$ and $G^{\prime}$ from \autoref{ex:eq} and assume that, starting with a decision diagram representing the identity, for every gate applied from~$G$ all gates from $G^{\prime}$ until the next red barrier shown in \autoref{fig:qftdecomp} are applied. 
	Applying the gates from~$G$ and~$G^\prime$ in such a particular order ``from the left'' and ``from the right'', respectively,
	yields situations where the impact of a gate from circuit~$G$  (increasing the size of the decision diagram) is reverted by multiplications with inverted gates from $G^{\prime}$ (decreasing the size of the decision diagram back to the representation of the identity function). 
	This way, the equivalence check can be conducted on much smaller intermediate representations and, hence, much more efficiently.
\end{example}

Moreover, even if the considered circuits~$G$ and~$G^\prime$ are \emph{not} functionally equivalent (and, hence, identity is not achieved), the observations from above still promise improvements compared to creating the complete decision diagrams for~$G$ and~$G^\prime$. This is, because in this case, %
the result of $G\shortrightarrow \mathbb{I}\shortleftarrow G^{\prime}$ inherently provides an efficient representation of the circuit's difference that allows one to obtain %
counterexamples almost ``for free'' (while those have to be explicitly generated using additional inversion and multiplication operations otherwise).

Overall, following those ideas, equivalence checking of two quantum circuits can be conducted very efficiently on rather compact decision diagrams, as shown in~\cite{burgholzerAdvancedEquivalenceChecking2021}.
But determining when to apply gates from~$G$ and when to apply (inverted) gates from~$G^{\prime}$ is not at all obvious.
Designing dedicated strategies for specific applications is a topic of ongoing research. 
As an example, a dedicated strategy for verifying the results of compilation flows can be derived by exploiting knowledge about the compilation flow itself~\cite{burgholzerVerifyingResultsIBM2020}.
An implementation of this method is available
at \url{https://github.com/cda-tum/qcec} and via the corresponding Python package \href{https://pypi.org/project/mqt.qcec/}{\texttt{mqt.qcec}}.
In addition, \url{https://www.cda.cit.tum.de/app/ddvis/} provides an installation-free visualization scheme of the procedure that also can be used to try out different gate-application schemes.

\subsection{The Power of Simulation}\label{sec:powerofsim}
The second characteristic we are exploiting rests on the observation that simulation is much more powerful for equivalence checking of quantum circuits than for equivalence checking of classical circuits.
More precisely, in the classical realm, it is certainly possible to simulate two circuits with random inputs to obtain counterexamples in case they are not equivalent. 
However, this often does not yield the desired result. In fact, due to masking effects and the inevitable information loss introduced by many classical gates, the chance of detecting differences in the circuits within a few arbitrary simulations is greatly reduced (e.g., $x \wedge 0$ masks any difference that potentially occurs during the calculation of~$x$).
Consequently, sophisticated schemes for constraint-based stimuli generation~\cite{yuanConstraintbasedVerification2006,bergeronWritingTestbenchesUsing2006,kitchenStimulusGenerationConstrained2007,willeSMTbasedStimuliGeneration2009}, fuzzing~\cite{laeuferRFUZZCoveragedirectedFuzz2018,leDetectionHardwareTrojans2019}, etc. are employed in order to verify classical circuits. 

In quantum computing, the inherent reversibility of quantum operations dramatically reduces these effects and frequently yields situations where even small differences remain unmasked and affect entire system matrices---showing the power %
of random simulations for checking the equivalence of quantum circuits. %
Because of that, it is in general not necessary to compare the \emph{entire} system matrices---in particular when two circuits are \emph{not} equivalent and, hence, their system matrices differ from each other.

Given two unitary matrices $U$ and $U^\prime$, we define their \emph{difference} $D$ as the unitary matrix $D=U^{\dag}U^\prime$ and it holds that $U \cdot D = U^\prime$.
In case both matrices are identical (i.e., the circuits are equivalent), it directly follows that \mbox{$D=\mathbb{I}$}.
One characteristic of the identity function~$\mathbb{I}$ resulting in this case is that all diagonal entries are equal to one, i.e.,~$\bra{i} U^{\dag}U^\prime \ket{i} = 1$ for $i\in\{0,\dots,2^n-1\}$, where $\ket{i}$ denotes the $i^{th}$ computational basis state.
More generally---in case of a potential relative/global phase difference between $G$ and $G^\prime$---all diagonal elements have modulus one, i.e.,~\mbox{$\vert\bra{i} U^{\dag}U^\prime \ket{i}\vert^2 = 1$}.
This expression can further be rewritten to %
\begin{align*}
1 &= \vert\bra{i} U^{\dag}U^\prime \ket{i}\vert^2 = \vert(U\ket{i})^\dag (U^\prime \ket{i})\vert^2
 = \vert\ket{u_i}^\dag \ket{u_i^\prime}\vert^2 \\&= \vert\braket{u_i}{u_i^\prime}\vert^2,
\end{align*}
where $\ket{u_i}$ and $\ket{u_i^\prime}$ denote the $i^{th}$ column of $U$ and $U^\prime$, respectively.
This essentially resembles the simulation of both circuits with the initial state $\ket{i}$ and, afterwards, calculating the fidelity $\mathcal{F}$ between the resulting states $\ket{u_i}$ and $\ket{u_i^\prime}$.  
Hence, if only one simulation yields~\mbox{$\mathcal{F}_i\coloneqq\mathcal{F}(\ket{u_i}, \ket{u_i^\prime}) \not\approx 1$}, then $\ket{i}$ proves the \mbox{non-equivalence} of  $G$ and $G^\prime$.

This constitutes an exponentially easier task than constructing the entire system matrices $U$ and $U^\prime$---although the complexity of simulation still remains exponential with respect to the number of qubits (for which the decision diagram-based solution reviewed above provides an efficient solution in many cases).
Regarding the complexity, creating the entire system matrices corresponds to simulating the respective circuit with all $2^n$ different computational basis states.
All this, of course, does not guarantee that any difference is indeed detected by just simulating a limited number of arbitrary computational basis states~$\ket{i}$.
This brings up the following question: How significantly do the matrices $U$ and $U^\prime$ differ from each other in case of \mbox{non-equivalence}, i.e., how many computational basis states~$\ket{i}$ yield $\mathcal{F}_i \not\approx 1$ for a given difference matrix $D$.
Since the difference $D$ of both matrices is unitary itself, it may as well be interpreted as a quantum circuit $G_D$. 
In the following, we assume that each gate of $G_D$ either represents a single-qubit or a multi-controlled operation\footnote{This does not limit the applicability of the following findings, since arbitrary single-qubit operations combined with {\sc CNOT} form a universal gate-set~\cite{nielsenQuantumComputationQuantum2010}.}.

\begin{example}\label{ex:bestcase}
	Assume that $G_D$ only consists of one (non-trivial) single-qubit operation defined by the matrix $U_s$ applied to the first of $n$ qubits. Then, the system matrix $D$ is given by $\mathit{diag}(U_s,\dots, U_s)$.
	The process of going from $U$ to $U^\prime$, i.e., calculating $U\cdot D$, impacts \emph{all} columns of $U$.
	Thus, an error is detected by a \emph{single} simulation with \emph{any} computational basis state.
\end{example}

Among all quantum operations, single-qubit operations possess a system matrix least similar to the identity matrix due to the tensor product structure of their corresponding system matrix. 

\begin{example}\label{ex:worstcase}
	In contrast to \autoref{ex:bestcase}, assume that $G_D$ only consists of one operation $U_s$ targeted at the first qubit and controlled by the remaining $n-1$ qubits. Then, the corresponding system matrix is given by $\mathit{diag}(\mathbb{I}_2,\dots, \mathbb{I}_2, U_s)$.
	In this case, applying $D$ to $U$ only affects the last two columns of $U$. As a consequence, a maximum of two columns (out of~$2^n$) may serve as counterexamples---the worst case scenario.
\end{example}

These basic examples cover the extreme cases when it comes to the difference of two unitary matrices. In case $G_D$ exhibits no such simple structure, the analysis is more involved, e.g., generally quantum operations with \mbox{$c\in\{0,\dots,n-1\}$} controls will exhibit a difference in $2^{n-c}$~columns. Furthermore, given two operations showing a certain number of differences, the matrix product of these operations in most cases (except when cancellations occur) differs in as many columns as the maximum of both operands.

The gate-set provided by (current) quantum computers typically includes only (certain) single-qubit gates and a specific two qubit gate, such as the \qop{CNOT}~gate. Thus, multi-controlled quantum operations usually only arise at the most abstract algorithmic description of a quantum circuit and are then \emph{decomposed} into elementary operations from the device's \mbox{gate-set} before the circuit is mapped to the target architecture.
As a consequence, errors occurring during the design flow will typically consist of (1)~single-qubit errors, e.g., offsets in the rotation angle, or (2)~errors related to the application of \qop{CNOT} or \qop{SWAP}~gates.
In both cases, non-equivalence can be efficiently concluded by a limited number of simulations with arbitrary computational basis states.
If a counterexample was not obtained after a few simulations, this yields a highly probable estimate of the circuit's equivalence---in contrast to the classical realm, where this generally does not allow for any conclusion.
Further details and evaluations on the respective methods are available in~\cite{burgholzerRandomStimuliGeneration2021, burgholzerAdvancedEquivalenceChecking2021}.

\section{Conclusions}\label{sec:conclusions}

The power of quantum computing comes with new computing primitives and the need for suitable design automation methods.
In this work, we reviewed decision diagrams for quantum computing as well as  
their application in quantum circuit simulation (with and without noise) as well as the verification of quantum circuits.
Decision diagrams offer a complementary approach for tackling the complexity of these tasks with a potential impact comparable to their conventional counterparts. With this work, we want to encourage their usage in the quantum (design automation) community. Implementations of the methods presented here are available at the corresponding GitHub repositories mentioned above.

\section*{Acknowledgments}

We sincerely thank all co-authors and collaborators who work(ed) with us in this exciting area. 
Special thanks go to Alwin Zulehner and Thomas Grurl.

This work received funding from the European Research Council (ERC) under the European Union’s Horizon 2020 research and innovation program (grant agreement No. 101001318), was part of the Munich Quantum Valley, which is supported by the Bavarian state government with funds from the Hightech Agenda Bayern Plus, and has been supported by the BMWK on the basis of a decision by the German Bundestag through project QuaST, as well as by the BMK, BMDW, and the State of Upper Austria in the frame of the COMET program (managed by the FFG).
\printbibliography

\end{document}